\documentclass[12pt,preprint]{aastex}

\shorttitle{IRAC OBSERVATIONS OF WHITE DWARFS. II.}
\shortauthors{FARIHI, BECKLIN, \& ZUCKERMAN}

\begin{document}

\title{{\em SPITZER} IRAC OBSERVATIONS OF WHITE DWARFS. II.  
		MASSIVE PLANETARY AND COLD BROWN DWARF
		COMPANIONS TO YOUNG AND OLD DEGENERATES}

\author{J. Farihi\altaffilmark{1,2},
	 E. E. Becklin\altaffilmark{1}, \& 
	 B. Zuckerman\altaffilmark{1}}

\altaffiltext{1}{Department of Physics \& Astronomy,
			University of California,
			430 Portola Plaza,
			Los Angeles, CA 90095; jfarihi,becklin,ben@astro.ucla.edu}
\altaffiltext{2}{Gemini Observatory,
			Northern Operations,
			670 North A'ohoku Place,
			Hilo, HI 96720}

\begin{abstract}

This paper presents a sensitive and comprehensive IRAC $3-8$ $\mu$m photometric 
survey of white dwarfs for companions in the planetary mass regime with temperatures 
cooler than the known T dwarfs.  The search focuses on descendents of intermediate 
mass stars with $M\ga3$ $M_{\odot}$ whose inner, few hundred AU regions cannot 
be probed effectively for massive planets and brown dwarfs by any alternative existing 
method.  Furthermore, examination for mid-infrared excess explores an extensive range 
of orbital semimajor axes, including the intermediate $5-50$ AU range poorly covered 
and incompletely accessible by other techniques at main sequence or evolved stars.
Three samples of white dwarfs are chosen which together represent relatively young as 
well as older populations of stars: 9 open cluster white dwarfs, 22 high mass field white 
dwarfs, and 17 metal-rich field white dwarfs.  In particular, these targets include: 7 Hyads 
and 4 field white dwarfs of similar age; 1 Pleiad and 19 field white dwarfs of similar age; 
van Maanen 2 and 16 similarly metal-rich white dwarfs with ages between 1 and 7 Gyr.  

No substellar companion candidates were identified at any star.  By demanding a 15\% 
minimum photometric excess at 4.5 $\mu$m to indicate a companion detection, upper limits 
in the planetary mass regime are established at 34 of the sample white dwarfs, 20 of which 
have limits below 10 $M_{\rm J}$ according to substellar cooling models.  Specifically, limits
below the minimum mass for deuterium burning are established at all Pleiades and Hyades
white dwarfs, as well as similarly young field white dwarfs, half a dozen of which receive limits 
at or below 5 $M_{\rm J}$.  Two IRAC epochs at vMa 2 rule out $T\ga200$ K proper motion 
companions within 1200 AU.

\end{abstract}


\section{INTRODUCTION}

The first strong candidate and certain substellar objects identified outside the Solar system 
were all discovered orbiting evolved degenerate stars: the probable brown dwarf companion to
the white dwarf GD 165 \citep*{kir99,bec88}, and the planetary system orbiting the pulsar PSR 
1257$+$12 \citep*{wol92}.  Furthermore, the first directly detected, unambiguous substellar and 
planetary mass objects were imaged as wide companions orbiting low luminosity primaries: the 
brown dwarf secondary to the M dwarf Gl 229 \citep*{nak95}, and the planetary mass secondary 
to the young brown dwarf commonly known as 2M1207 \citep*{son06,cha05}.  Historically, as 
well as astrophysically, these properties provide a clear advantage over other types of primaries 
in the quest to directly detect radiation from bound substellar objects of the lowest mass, such 
as planets.

Observations indicate that bound substellar objects and planetary system components survive 
post-main sequence evolution.  First, there now exist roughly one dozen first ascent giant stars 
known to harbor substellar and planetary companions \citep*{nie07,ref06,hat06,hat05,sat03,fri02}.
Second, there are at least 3 white dwarfs which have close, unaltered and unevolved substellar 
companions \citep*{bur06,max06,far05a}.  Third, there are at least 10 white dwarfs with infrared
excess due to debris disks which indicate a growing probability of orbiting rocky planetesimals 
\citep*{far08,jur07b,jur07a,kil07,rea05}.  These cool white dwarfs with warm orbiting dust also 
display anomalous photospheric metals which are almost certainly accreted from their circumstellar 
material.  The origin of the orbiting material and the dynamical interactions necessary to bring it close 
enough to the star to be accreted, are consistent with remnant planetary systems \citep*{jur03,deb02}.

The earth-size radii and consequent low luminosities typical of white dwarfs are clear advantages 
when searching for light emitted from cold jupiter-size planets and brown dwarfs.  Prior to the 
launch of {\em Spitzer}, generally speaking, only M and L dwarf companions could be detected 
directly as excess infrared emission from white dwarfs \citep*{tre07,hoa07,bur06,far06,far05a,
far05b,far04a,wac03,zuc87,pro83}.  As a benchmark, a typical 10000 K degenerate and an 
L5 dwarf are about equally luminous at $K$-band \citep*{dah02,ber95b}, while ground-based 
observations of white dwarfs at longer wavelengths (where the contrast for cooler companions 
would improve) are prohibited by overwhelming sky brightness \citep*{gla99}.

Owing to the capabilities of {\em Spitzer} \citep*{wer04}, a Cycle 1 IRAC \citep*{faz04} program 
was undertaken to photometrically search for massive planets and cold substellar companions 
to relatively young and old white dwarfs, respectively.  Specifically, the target sample includes
white dwarfs in the Hyades and Pleiades, high mass field white dwarfs, and metal-rich field
white dwarfs.  Farihi, Zuckerman, \& Becklin (2008; Paper I) describe photometry for all the 
older (metal-rich) degenerate targets, while this paper presents a synopsis of the IRAC results 
for the younger (open cluster and high mass) degenerates, and upper mass limits for unresolved 
companions to all the Cycle 1 targets.

\section{RELATIVELY YOUNG AND OLD DEGENERATE TARGETS}

While highly evolved, white dwarfs are not necessarily old.  This fact is exemplified by the 
nearest and brightest degenerate star in the sky, Sirius B, with a mass of $M\approx1.00$ 
$M_{\odot}$ and a total age of $\tau\approx240$ Myr \citep*{lie05b}.  For stars which evolved 
essentially as single objects, there is a correlation between main sequence progenitor mass and 
white dwarf mass, derived primarily from studies of open clusters \citep*{dob06a,kal05,wil04,cla01,
wei00,wei90,wei87}.  This initial-to-final mass relation yields an estimate of the total age for any 
particular white dwarf provided its mass and cooling age are accurately known.  For white dwarf 
masses below 0.6 $M_{\odot}$, the initial-to-final mass relation is quite steep and small errors in 
degenerate mass can lead to large errors in main sequence lifetime \citep*{bur02}.

\subsection{Hyad and Pleiad Targets}

The Hyades and Pleiades are relatively young and nearby open clusters; the former at $d=46$ 
pc and $\tau=625$ Myr \citep*{per98,pin98}, and the latter at $\tau=125$ Myr and $d=132$ pc 
\citep*{sta98,pin98,sod98}.  Classically, the only white dwarf Pleiad is EG 25 (LB 1497), but recently 
the possibility has been raised that the massive white dwarfs GD 50 and PG 0136$+$251 may have 
originated in the same region which gave rise to the cluster \citep*{dob06b}.  The latter stars are 
considered in the following section, and only EG 25 is listed among cluster targets.  Table \ref{tbl1} 
lists all observed open cluster white dwarfs, including the seven classical single white dwarf Hyads 
and EG 265 (V411 $\tau$), which is either a proper cluster member or part of the Hyades supercluster 
\citep*{rei93,egg84}.  Excluded are the white dwarf Hyads in binary systems; V471 $\tau$ and HZ 9.  
The open cluster targets come from this study ({\em Spitzer} Program 3549; PI Becklin), with the exception 
of Hyades targets EG 39 and EG 42, which were extracted from the {\em Spitzer} archive (Program 2313; 
PI Kuchner).  

\subsection{High Mass Targets}

As with Sirius B and the white dwarf Pleiad, a white dwarf mass near 1.0 $M_{\odot}$ implies 
a short main sequence lifetime for unadulterated single star evolution, regardless of association.  
More specifically, were these young white dwarfs identical in mass (depending on the reference, 
their masses differ by no more than 10\%; \citealt*{lie05b,cla01}), their total age difference can be
estimated from the difference in their effective temperatures; $T_{\rm eff}=25,200$ K for Sirius B, and 
$T_{\rm eff}=31,700$ K for EG 25.  For these temperatures, log $g=8.6$ (very nearly 1.0 $M_{\odot}$) 
hydrogen atmosphere models predict cooling ages of 130 and 60 Myr respectively, which by itself 
would account for 60\% of their total age difference \citep*{bar05,cla01,ber95b,ber95c,ber95a}.  
In reality the total difference in their ages is due both to differential cooling and unequal main 
sequence progenitor lifetimes.

Any hot, high mass white dwarf which evolved as a single star will be similarly young, or even 
younger for higher temperatures or masses.  In this paper, it is assumed that all high mass white 
dwarfs are descended from single main sequence star progenitors of intermediate mass, but this may 
not be the case.  There appears to be indirect evidence in favor of, as well as against, the existence of 
high mass white dwarfs resulting from mergers, but no direct evidence exists \citep*{han06,fer05,lie05a}.
Ten hot and massive white dwarf targets come from this study ({\em Spitzer} Program 3549; PI Becklin) 
and one dozen similar targets were extracted from the {\em Spitzer} archive (Program 3309; PI Hansen).  
Table \ref{tbl2} lists the 22 hot field white dwarfs with masses $M\geq0.9$ $M_{\odot}$ selected for study.  

\subsection{Older Targets}

Included in the sample are metal-rich white dwarfs from Paper I ({\em Spitzer} Program 3548; PI Zuckerman) 
with the addition of vMa 2, which was extracted from the {\em Spitzer} archive (Program 33; PI Fazio).  
These targets are relatively old, with total ages between $1-7$ Gyr and are listed in Table \ref{tbl3} 
for completeness, although their IRAC fluxes are previously published in Paper I with the exception 
of vMa 2.
 
\section{OBSERVATIONS AND DATA}

For white dwarf targets in Programs 3548 and 3549, the details of the IRAC observing strategy, 
data reduction and analysis are described in full detail in Paper I.  In these programs a total 
exposure time of 600 s was utilized for each target in all bandpasses.  For white dwarf targets 
extracted from the {\em Spitzer} archive, the exposure times were shorter and occasionally in 
only 2 bandpasses (see \citealt*{mul07,han06}).  Fortunately, all targets were unambiguously 
detected at 4.5 $\mu$m, the wavelength which places the best constraints on spatially unresolved 
cold substellar and massive planetary companions, according to models for the appropriate range 
of ages and masses \citep*{bur03,bar03}.

\subsection{Photometry and Upper Limits}

Paper I contains a detailed discussion regarding the consistency of measured IRAC fluxes of white 
dwarfs compared to model predictions and concludes the photometric accuracy is well described 
as 5\%.  Also described there is a conservative approach to estimating the signal-to-noise of IRAC 
detections in the presence of possible confusion and spatially varying background.  In Paper I, all 
targets were detected at all wavelengths, which is not the case here.

To create upper limits for nondetections, aperture photometry was performed as described 
in Paper I at the nominal location of the white dwarf, derived from one or more IRAC channels
in which the source was positively detected.  Utilizing the smallest radius ($r=2$ pixels) for which 
there are published aperture corrections, the flux in this aperture was compared to the per pixel 
sky noise multiplied by the area of the aperture, and the larger of these values was taken to be 
the upper limit, after an appropriate aperture correction.  In nearly all cases, the larger value 
was given by the additive noise in the aperture, but there were a few cases in which there was 
flux measured above this level.  In these cases, it appeared possible or likely that the measured
flux originated from background sources as evidenced by pixel shifts in the centroid of the source 
flux compared to the other IRAC channels, or sources which were apparently extended.

\begin{figure}
\plotone{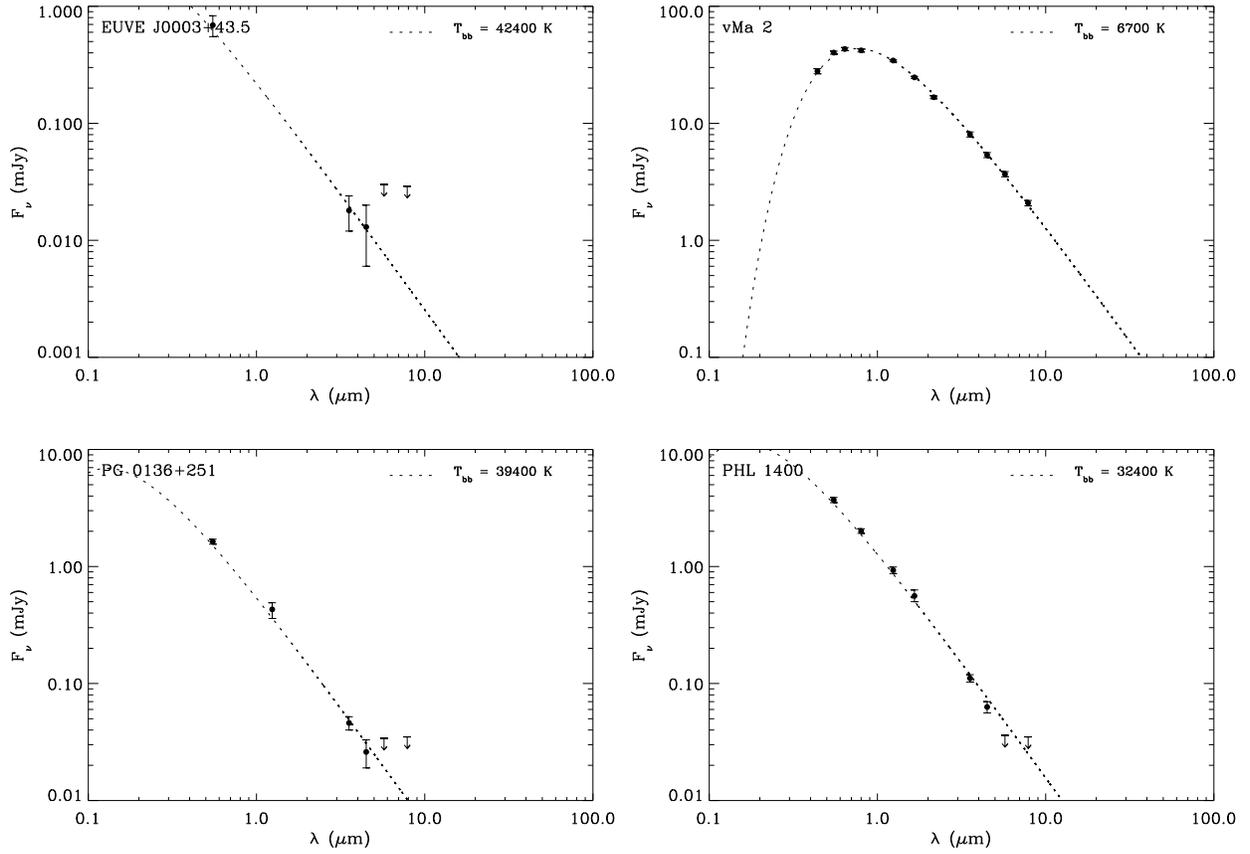}
\caption{Spectral energy distribution of EUVE J0003$+$43.5, vMa 2, PG 0136$+$251, and
PHL 1400.  Downward arrows represent upper limits (\S3.1).
\label{fig1}}
\end{figure}

\begin{figure}
\plotone{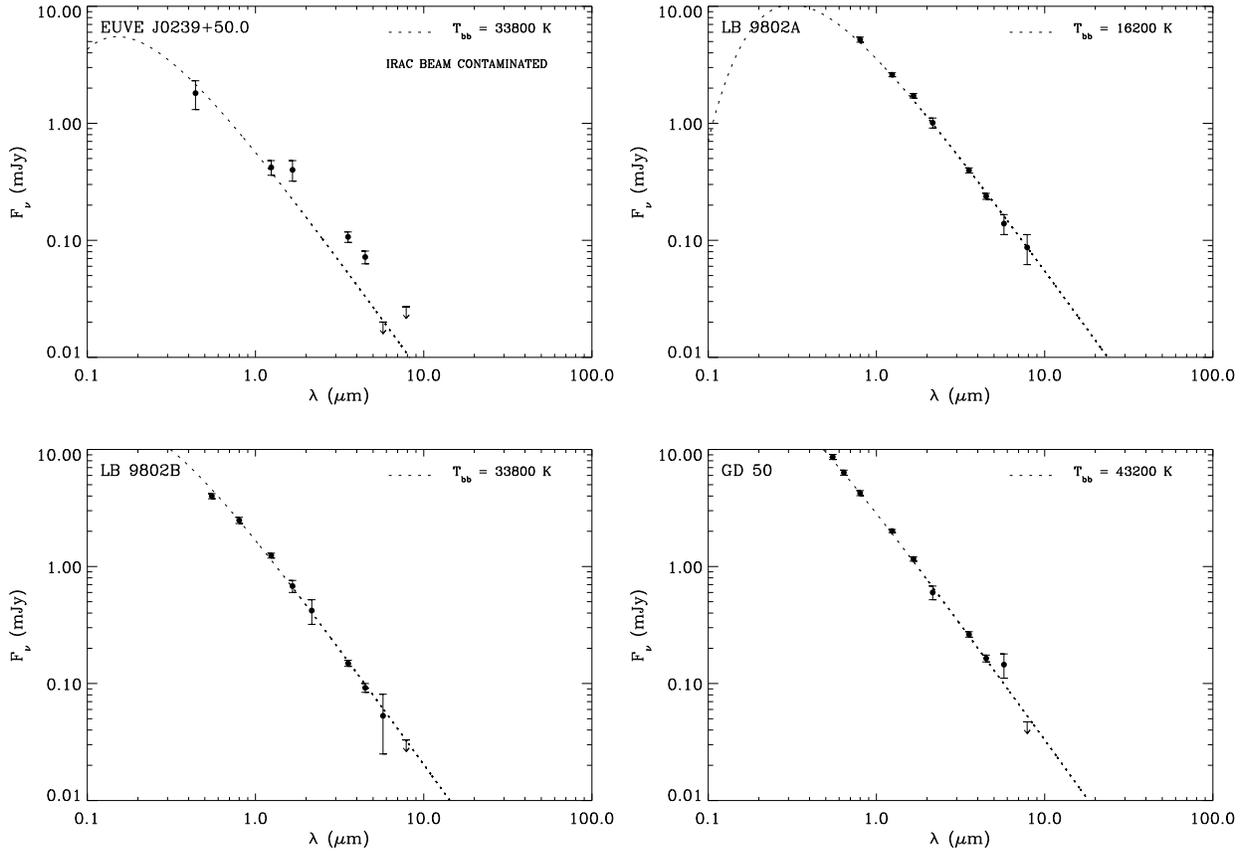}
\caption{Spectral energy distribution of EUVE J0239$+$50.0, LB 9802A, LB 9802B, and
GD 50.  Downward arrows represent upper limits (\S3.1).
\label{fig2}}
\end{figure}

\begin{figure}
\plotone{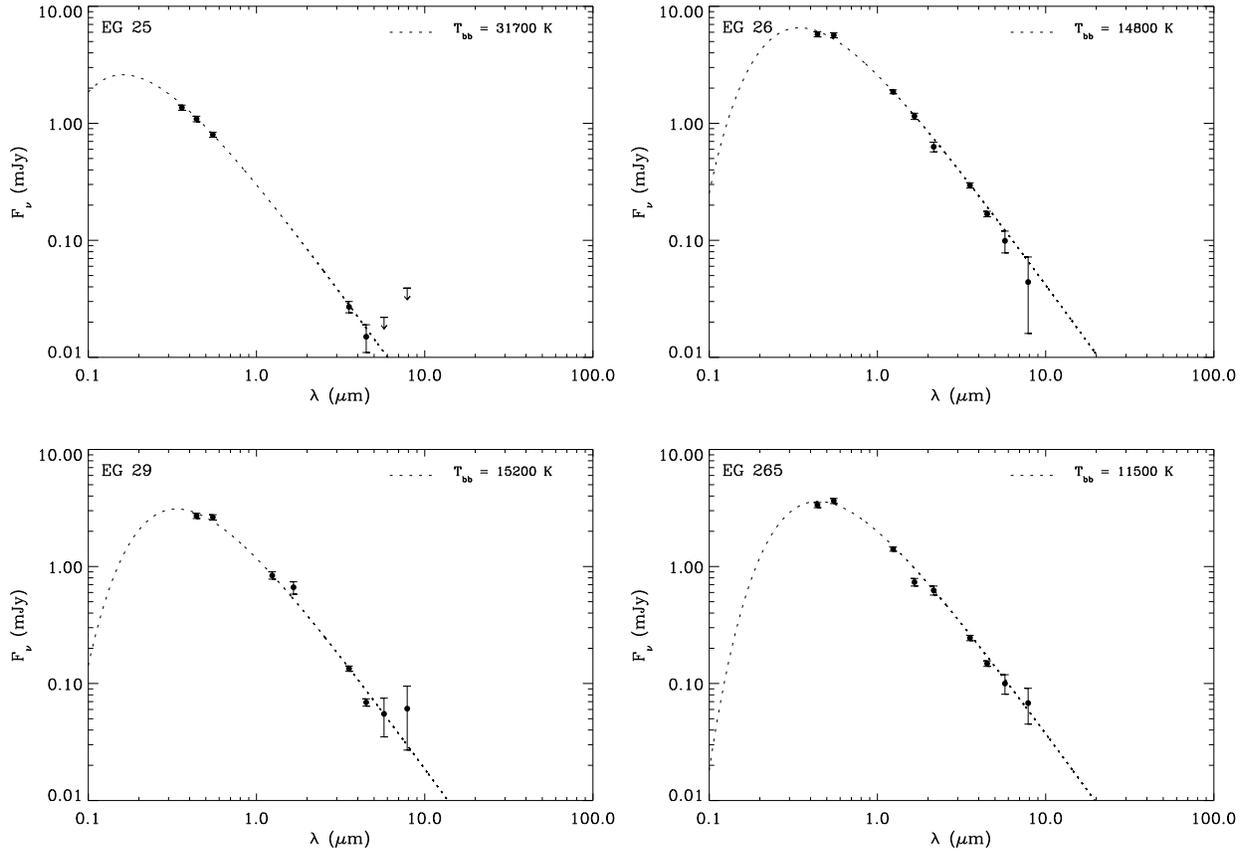}
\caption{Spectral energy distribution of EG 25, EG 26, EG 29, and EG 265.  Downward arrows 
represent upper limits (\S3.1).
\label{fig3}}
\end{figure}

\begin{figure}
\plotone{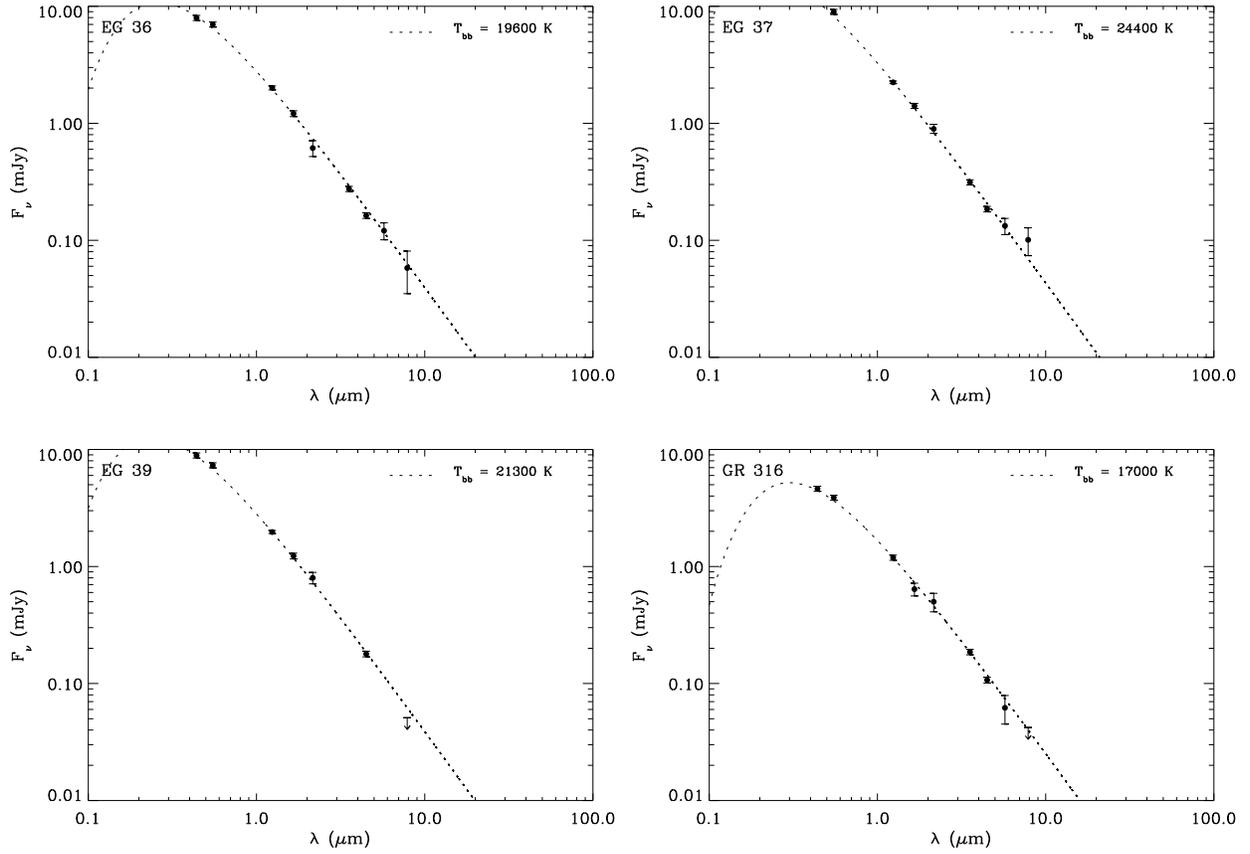}
\caption{Spectral energy distribution of EG 35, EG 37, EG 39, and GR 316.  Downward arrows 
represent upper limits (\S3.1).
\label{fig4}}
\end{figure}

\begin{figure}
\plotone{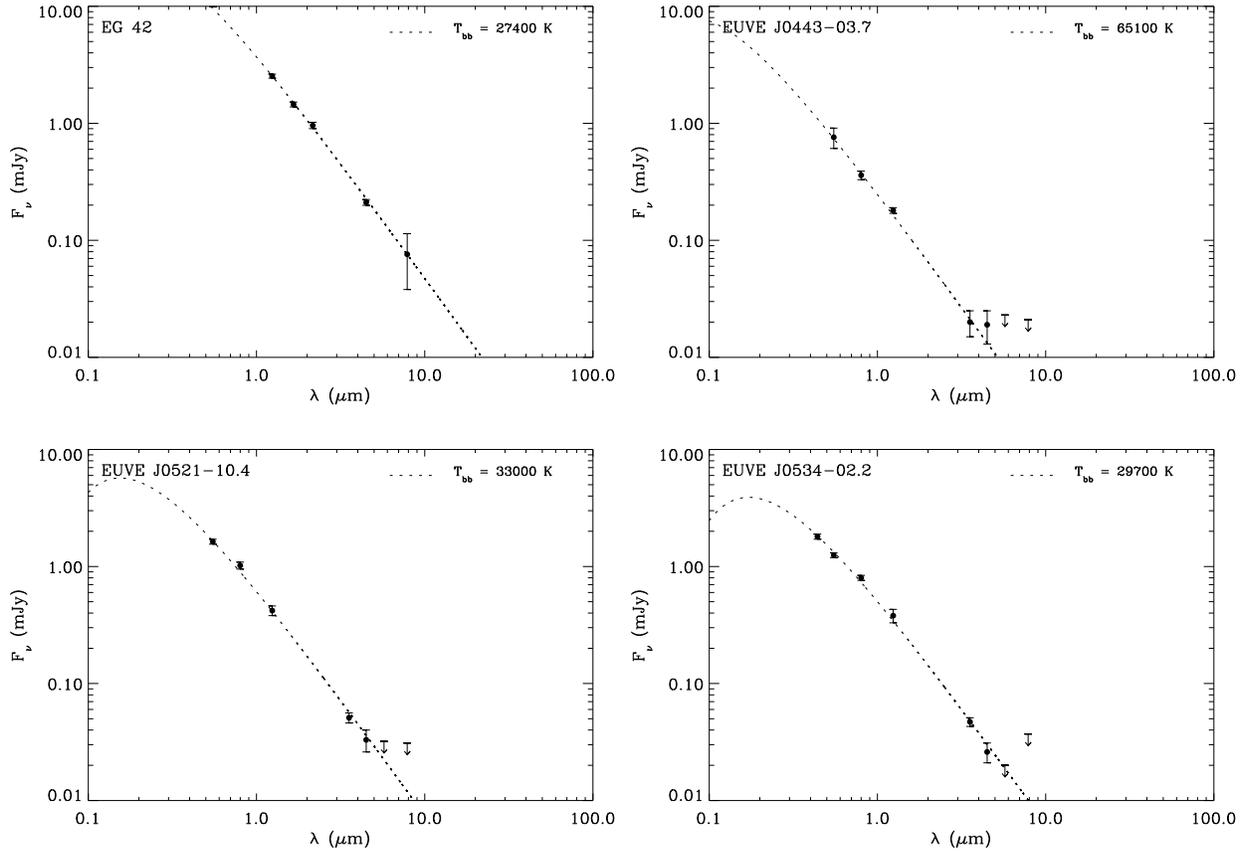}
\caption{Spectral energy distribution of EG 42, EUVE J0443$-$03.7, EUVE J0521$-$10.4, 
and EUVE J0534$-$02.2.  Downward arrows represent upper limits (\S3.1).
\label{fig5}}
\end{figure}

\begin{figure}
\plotone{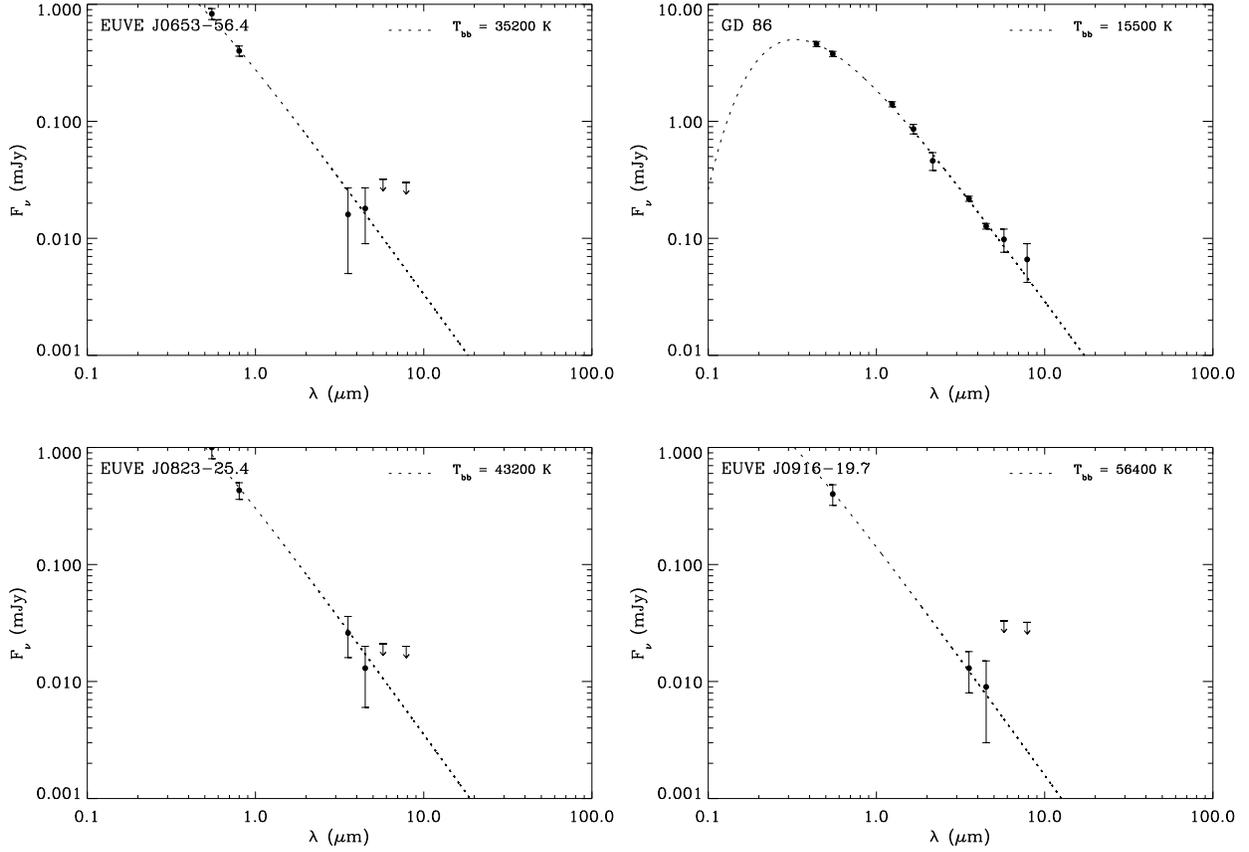}
\caption{Spectral energy distribution of EUVE J0653$-$56.4, GD 86, EUVE J0823$-$25.4,
and EUVE J0916$-$19.7.  Downward arrows represent upper limits (\S3.1).
\label{fig6}}
\end{figure}

\begin{figure}
\plotone{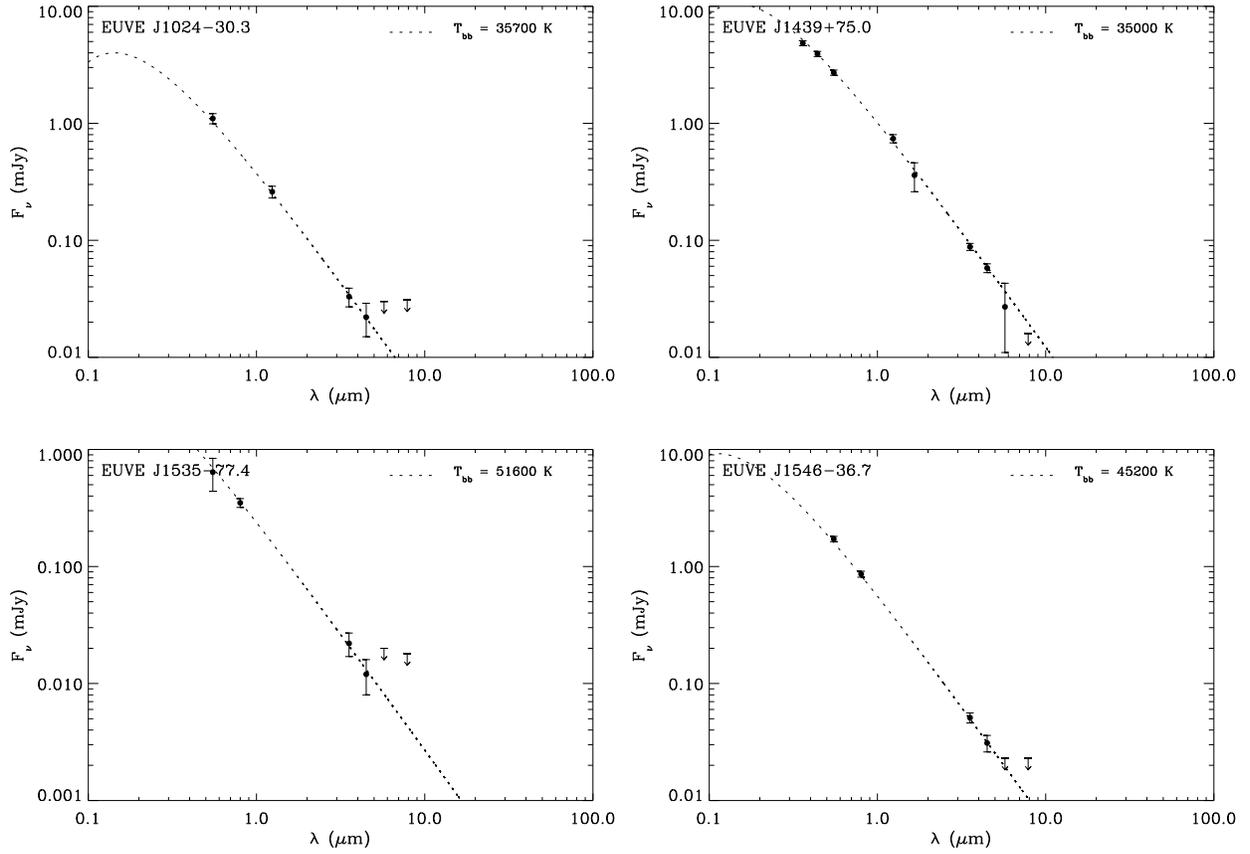}
\caption{Spectral energy distribution of EUVE J1024$-$30.3, EUVE J1439$+$75.0, 
EUVE J1535$-$77.4, and EUVE J1546$-$36.7.  Downward arrows represent upper limits (\S3.1).
\label{fig7}}
\end{figure}

\begin{figure}
\plotone{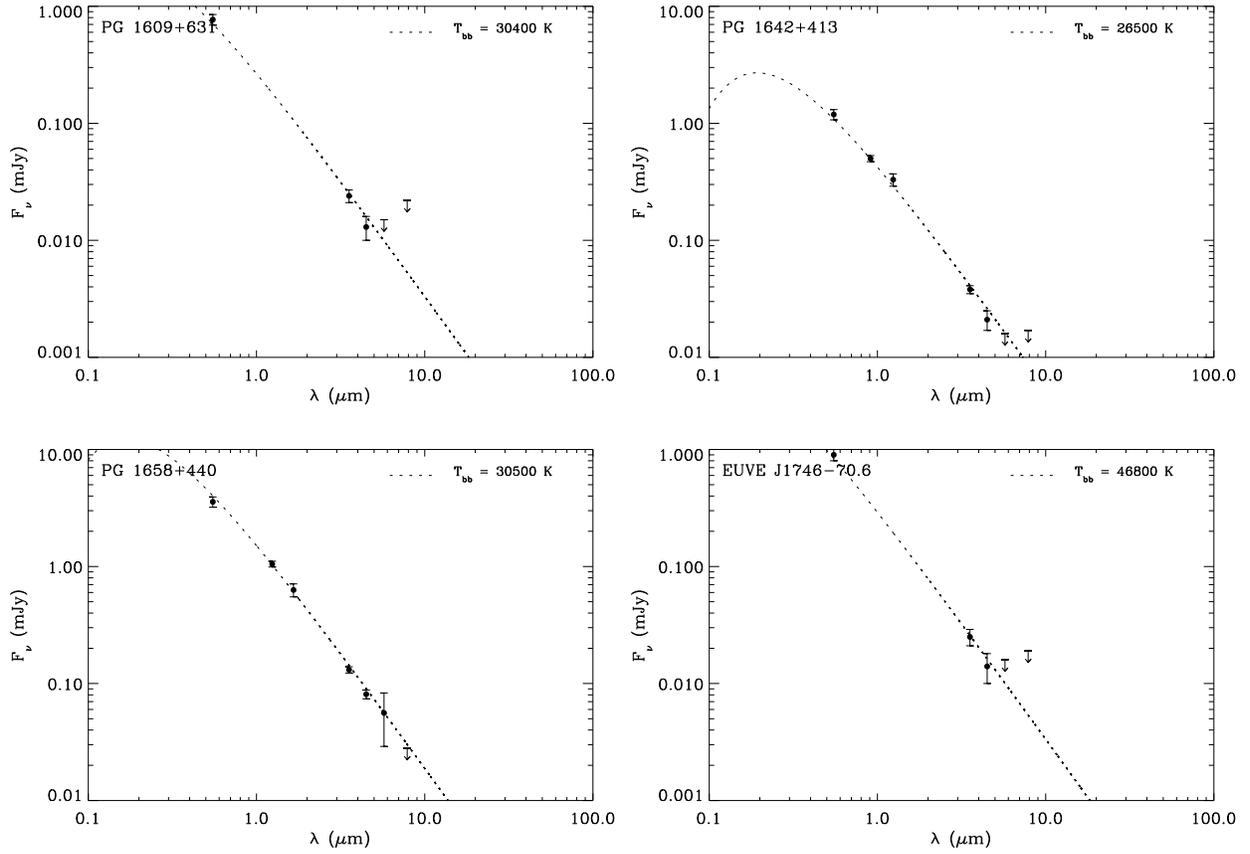}
\caption{Spectral energy distribution of PG 1609+631, PG 1642$+$413, PG 1658$+$440,
and EUVE J1746$-$70.6.  Downward arrows represent upper limits (\S3.1).
\label{fig8}}
\end{figure}

It is noteworthy that the photometric errors used here, as well as the derived upper limits for 
nondetections, are somewhat larger than those published in \citet*{han06} and \citet*{mul07} 
for the same observations.  In some cases where white dwarf flux is reported by those authors, 
Table \ref{tbl4} indicates only an upper limit for the reasons stated at the end of the previous 
paragraph.  IRAC fluxes and upper limits for 32 of 48 studied stars (excepting those previously 
published in Paper I) are listed in Table \ref{tbl4} and plotted in Figures \ref{fig1}--\ref{fig8}.

\section{ANALYSIS}

\subsection{Total Ages}

To infer companion mass limits using substellar cooling models, each white dwarf 
target requires an assessment of its total age since formation as a main sequence star.  
Substellar companions which form in a binary will be truly coeval, while massive planets 
are thought to form within 10 Myr of their main sequence hosts.  For the open cluster white 
dwarfs, their total age is the cluster age, while for field white dwarfs the following methods 
are used to estimate their ages.

First, any field white dwarf with a mass and effective temperature similar to or higher 
than the Pleiad EG 25 is assumed to be of similar age, roughly 0.1 Gyr.  There are sufficient 
uncertainties in both the masses and temperatures of these stars, as evidenced by the $10-20$\% 
variation among parameters cited in the literature for the same objects (see Table \ref{tbl2} for a list 
of references) which translates into errors in cooling ages on the order of $10-20$ Myr.  Fortunately, 
this type of error should be offset when assessing a total age because higher mass white dwarfs cool 
more slowly (smaller surface area) yet have shorter inferred main sequence lifetimes than their less 
massive counterparts.  For the high mass field stars considered here, the modest errors in cooling age 
and inferred main sequence lifetime, which result from uncertainties in white dwarf parameters, are 
comparable in magnitude and therefore tend to cancel out.  Given these uncertainties for the hot and 
massive field white dwarfs, it seems prudent to assign a 20\% uncertainty in their total ages, or 
$\tau=0.125\pm0.025$ Gyr.

Second, all cooler and less massive (i.e. older) field degenerates have their total ages assessed 
following the procedure employed originally by \citet*{bur02} and more recently by \citet*{deb07}.
This latter method utilizes the initial-to-final mass to relation to obtain a main sequence mass from 
the current, known white dwarf mass.  A main sequence lifetime is then assigned based on the 
inferred main sequence mass, and this is added to the white dwarf cooling lifetime to obtain an 
approximate total age

\begin{equation}
\tau = t_{\rm ms} + t_{\rm wd}
\end{equation}

\noindent
Cooling ages come from models of P. Bergeron (2002, private communication; 
\citealt*{ber95b,ber95c}), while main sequence lifetimes were calculcated using the formulae of 
\citet*{hur00}.  Tables \ref{tbl5} and \ref{tbl6} list the relevant ages for all targets together with upper 
mass limits for substellar companions determined as described below.  It should be stated that this 
general method is the best available to estimate the total age of white dwarfs not belonging to open 
clusters or multiple systems from which another age constraint might be gleaned.  However, there 
are several sources of uncertainty in the estimation of total ages, including but not limited to: the 
slope of the initial-to-final mass relation, assumed main sequence lifetimes, and white dwarf model 
uncertainties; specifically, spectroscopic parameter fits and cooling ages \citep*{kal08,dob06a,fer05}.  
Owing to these facts, the uncertainty in the total ages of older field white dwarfs is taken to be 25\%.  

\subsection{Unresolved Companion Mass Limits from 4.5 $\mu$m Fluxes}

All white dwarfs in Tables \ref{tbl1}--\ref{tbl3} have measured IRAC fluxes at 4.5 $\mu$m, where 
cold ($T_{\rm eff}<1000$ K) substellar objects are predicted to be brightest \citep*{bar03,bur03}.  
This bandpass is best for placing limits on any spatially unresolved substellar companions.  The 
flux errors in Table \ref{tbl4} are $1\sigma$ values, but for reasons discussed in Paper I, and to be 
conservative, an unambiguous detection of excess at this wavelength is defined here as $3\sigma$ 
above the expected white dwarf flux.  By demanding this level of excess, blackbody models will 
suffice to predict the expected 4.5 $\mu$m photospheric flux of the white dwarf targets (Paper I).  
The absolute magnitude corresponding to the minimally reliable excess flux from each white dwarf 
target, is then given by

\begin{equation}
M_{4.5\mu{\rm m}} = -2.5 \log \left( \frac{3 \sigma}{F_0} \left( \frac{d}{10} \right)^2 \right) 
\end{equation}

\noindent
where $\sigma$ is the total flux error in Jy at 4.5 $\mu$m from Table \ref{tbl4}, $d$ is the distance 
to the white dwarf in pc, and $F_0=179.7$ Jy \citep*{ssc06a}.  Substellar cooling models updated 
to include fluxes in the IRAC bandpasses were used to transform the expected flux into a mass for 
a given age (I. Baraffe 2007, private communication; \citealt*{bar03}).  Some representative, model 
predicted values for $M_{4.5\mu{\rm m}}$ at the ages of interest are given in Table \ref{tbl7}.  It is 
noteworthy that this analysis rules out unresolved sub-T dwarf companions at 25 white dwarfs in 
Table \ref{tbl5}; the known T dwarf sequence ends near $M_{4.5\mu{\rm m}}=13.5$ mag and T8
\citep*{pat06}, and limits at these stars reach 13.6 $\leq M_{4.5\mu{\rm m}} \leq$ 15.2 mag (typically
$M_{4.5\mu{\rm m}}=14.3\pm0.5$ mag).

Additionally, any spatially resolved point sources detected within several arcseconds of the white 
dwarf were photometrically examined for the possibility of companionship via their IRAC colors and 
any available ground-based photometric or astrometric data.  Owing to the higher sensitivity of the 
2 short IRAC wavelengths, some resolved substellar objects may not be detected at the 2 long
IRAC wavelengths, resulting in a potential ambiguity for some visual companions.  Generally speaking, 
no candidate companions were identified in this manner, but a few possibilties are discussed in \S5.7.

\subsection{Resolved Companion Mass Limits from 7.9 $\mu$m Fluxes}

In order to detect T and sub-T dwarfs as spatially resolved companions, and to differentiate such 
objects from background point-like sources, they must be reliably detected at all 4 IRAC wavelengths.
The 2 long wavelength IRAC channels in particular, together with the 2 short wavelength channels, 
provide unique information which should eliminate the color degeneracy between cool brown dwarfs 
and red extragalactic (point-like) sources in the near-infrared and short wavelength IRAC channels alone 
\citep*{pat06}.  This fact appreciably limits any wide field IRAC search for T dwarf companions due to the 
lower sensitivity of the long wavelength IRAC channels \citep*{ssc06b}, as detailed below.  There are 7 
white dwarfs from Table \ref{tbl3} which met the necessary criteria for such a search: 1) IRAC imaging of 
their surrounding fields in all 4 channels and; 2) a distance within approximately 20 pc.  The entire known 
T dwarf sequence (down to T8) should be detected at all 4 wavelengths at these distances \citep*{pat06}. 

\begin{figure}
\plotone{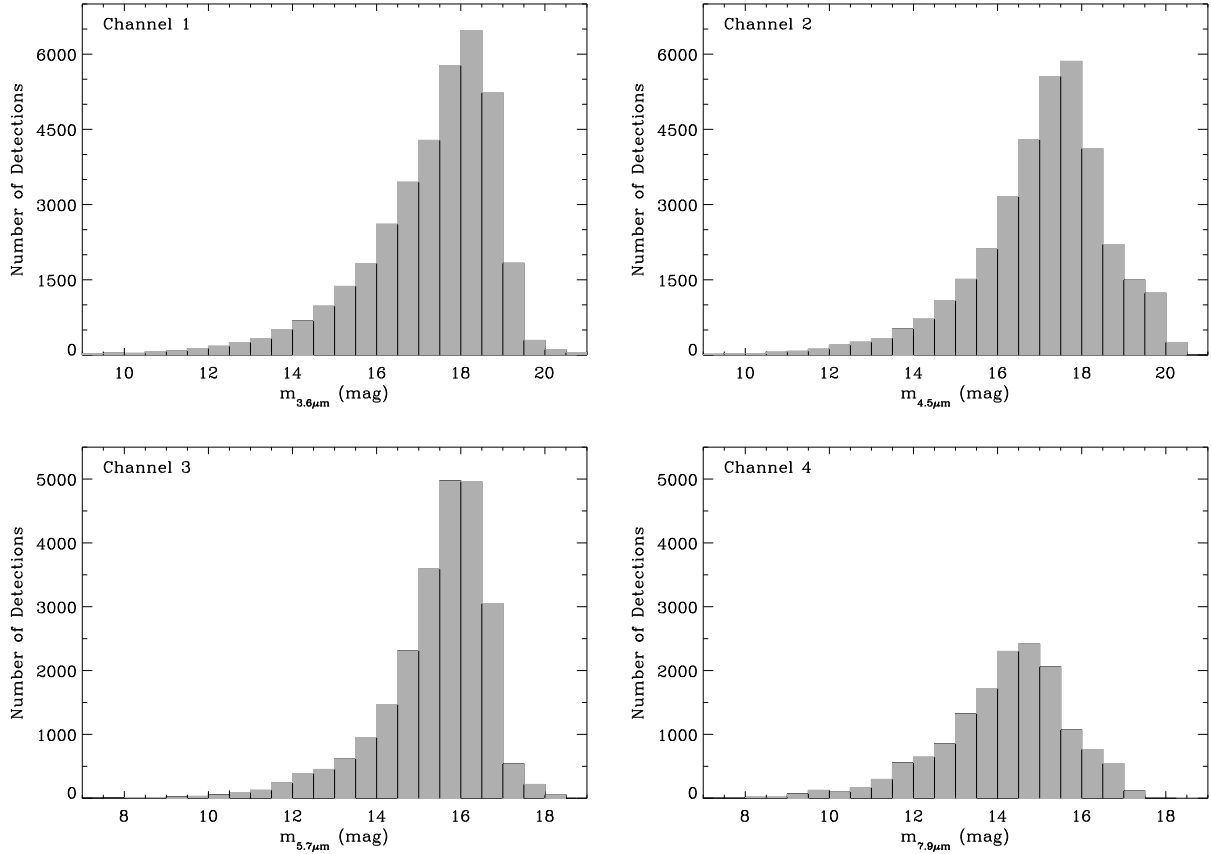}
\caption{The sum of the number of point-like source detections as a function of magnitude at each 
wavelength in the IRAC imaged fields of 16 metal-rich white dwarfs with 600 s exposure times (Paper I).
\label{fig9}}
\end{figure}

Figure \ref{fig9} shows the number of detected sources in each IRAC channel, as a function of 
magnitude, in the full IRAC fields of the 16 white dwarfs from Paper I which shared 600 s integration 
times per filter, and identical 20-point dithering patterns.  These sources were successfully detected 
and extracted photometrically by the IRAF tasks {\sf daofind} and {\sf daophot}.  Based on the number of 
detections per magnitude bin (disregarding any trends in the number of source counts as a function of 
wavelength), it is clear that the 7.9 $\mu$m channel would limit any 4 channel IRAC survey for objects 
whose spectral energy distributions are not rising towards longer wavelengths.

\subsubsection{Detectability of T and sub-T Dwarfs}

There exist a total of 58 white dwarfs which have been observed with IRAC at 7.9 $\mu$m utilizing 
a common experimental design; 22 targets from the present work, 16 white dwarfs from Paper I, and 
10 degenerates from \citet*{jur07a}.  This white dwarf dataset allows an empirical assessment of the
photometric sensitivity at this longest wavelength, and contains 35 unambiguous detections in that 
channel, with point sources as faint as 0.06 mJy reliably detected in all backgrounds.  This finding is 
consistent with: 1) the published sensitivities for IRAC \citep*{ssc06b}; 2) calculations by the Sensitivity 
Performance Estimation Tool; and 3) the number of sources detected as a function of magnitude in 
Figure \ref{fig9}.  Therefore at $m_{7.9\mu{\rm m}}=15.0$ mag, or 0.064 mJy, point sources should be 
well detected regardless of background.

The T dwarf sequence down to spectral type T8 ends at $M_{7.9\mu{\rm m}}=13.3$ mag 
\citep*{pat06}.  This corresponds to $m_{7.9\mu{\rm m}}=14.8$ mag at a distance of 20 pc and thus 
any widely separated T dwarf companions to the $d\leq20$ pc targets should be readily detected in 
this channel.  Similar calculations in the other 3 IRAC channels estimate that a T8 dwarf should 
be correspondingly well detected out to; 20 pc at 5.7 $\mu$m, 60 pc at 3.6 $\mu$m, and 100 pc at 4.5 
$\mu$m \citep*{pat06}.  For the white dwarf targets closer than 20 pc, the IRAC observations should be 
sensitive to small range of substellar companions with $M_{7.9\mu{\rm m}}>13.3$ mag, and potentially 
of a later spectral type than T.

Owing to the nature of the dithering pattern, a further assessment must be made in regards 
to the effective field of view for the depth described above.  For all but vMa 2, the medium sized 
cycling pattern was used, which should result in an effective coverage equal to the IRAC field of 
view minus about 25 pixels at each edge, or approximately $4.1'\times4.1'$, consistent with the 
analyzed images.  For vMa 2, the effective field of view is about 12 pixels larger at each edge, 
or approximately $4.6'\times4.6'$, and the sensitivity was 0.75 mag less at each channel owing 
to a 150 s total integration time.

\subsubsection{Selection of T and sub-T Dwarfs}

For each white dwarf searched for wide T and sub-T dwarf companions, all 4 IRAC filter 
images were aligned and combined to create a single master coordinate image. The IRAF 
task {\sf daofind} was executed on this master IRAC image to select point sources with counts 
at or above $3\sigma$ and the resulting coordinate list was then fed into {\sf daophot} for 
each filter image in order to perform automated point spread function fitting photometry. 

Template point spread functions were created by running {\sf daophot} on IRAC images of the 
Galactic component of the {\em Spitzer} Galactic First Look Survey to select approximately half 
to one dozen bright, unsaturated point sources in each filter.  The magnitudes of these selected 
template soures were calculated by creating zero points which included the zero magnitude fluxes 
for IRAC, multiplication by the appropriate unit area on the array, and the necessary conversion of 
units.  If the standard $r=10$ pixel radius aperture is used for photometry, these zero points are 
(17.30, 16.81, 16.33, 15.69) mag at (3.6, 4.5, 5.7, 7.9) $\mu$m.

The extracted sources in each filter were cross-correlated using the master coordinate list, and 
2 color-color diagrams were generated from the results; $m_{3.6\mu{\rm m}}-m_{4.5\mu{\rm m}}$ 
versus $m_{5.7\mu{\rm m}}-m_{7.9\mu{\rm m}}$, and $m_{3.6\mu{\rm m}}-m_{4.5\mu{\rm m}}$ versus 
$m_{4.5\mu{\rm m}}-m_{5.7\mu{\rm m}}$.  In these planes, T and sub-T dwarf candidates were selected 
by demanding an object meet 3 criteria suggested by the IRAC colors presented in \citet*{pat06}; 
$0.2< m_{3.6\mu{\rm m}}-m_{4.5\mu{\rm m}}<3.0$, $0.0<m_{5.7\mu{\rm m}}-m_{7.9\mu{\rm m}}<1.5$, 
and $-1.5<m_{4.5\mu{\rm m}}-m_{5.7\mu{\rm m}}<1.0$, with error bars ignored in the first cut.  All objects 
thus selected were examined individually; their images inspected and photometric data further evaluated.
Although the mid-infrared colors of sub-T type objects are somewhat uncertain, model predictions yield 
colors which fit with the selection criteria above (D. Saumon 2006, private communication).  Unfortunately, 
the earliest T dwarfs do not stand out strongly in IRAC color-color diagrams, and require near-infrared 
photometry to be clearly distinguished \citep*{pat06}.

No candidates which met all the criteria and passed critical examination were found in 
the IRAC fields of the 7 white dwarfs within $d=20$ pc.  While testing the color-color selection 
and extraction algorithm, the procedure was conducted at all 17 metal-rich white dwarfs in Table 
\ref{tbl3}.  A typical detection which met the color criteria had one or more of the following problems: 
1) large photometric errors; 2) a location near the noisy edge of the image; 3) probable confusion with 
another source; 4) association with a known image artifact; 5) a substantially discrepant color-magnitude 
at the expected distance.  Less often a detection would be a very red, unresolved extragalactic source 
whose nature was confirmed via existing optical astrometric and photometric catalogs. Table \ref{tbl6} 
lists the resulting upper limits on substellar mass companions achieved using this method, calculated 
by transforming $m_{7.9\mu{\rm m}}=15.0$ mag to the expected $M_{7.9\mu{\rm m}}$ at the target 
distance, then using substellar cooling models to obtain a mass from this flux at the appropriate total 
age (I. Baraffe 2007, private communication; \citealt*{bar03}).  

To test the selection and extraction algorithm, the IRAC fields of 3 T dwarfs were downloaded 
from the {\em Spitzer} archive (Program 35; PI Fazio) and the procedure was run on images containing 
objects with spectral types T2.0, T5.0, and T8.0 \citep*{pat06}.  All 3 objects were selected correctly 
by the color-color cuts, and with modest photometric error bars indicating genuine detections.  There 
is a non-zero probability that a bona fide cold brown dwarf companion escaped detection among our
white dwarf target fields, despite the ability of {\sf daophot} to spatially and photometrically deconvolve
overlapping point sources.  Based on the detection logs and the statistics from Figure \ref{fig9}, in the 
two short wavelength IRAC filter images there were roughly 1000 sources per field with brightnesses 
greater than the completeness limit in those bandpasses.  Taking a worst case scenario where all
these sources represent potential spoilers yields a 2\% probability of chance alignment within the 
relatively large, reduced image field of view.

\subsubsection{The Hyades}

\begin{figure}
\plotone{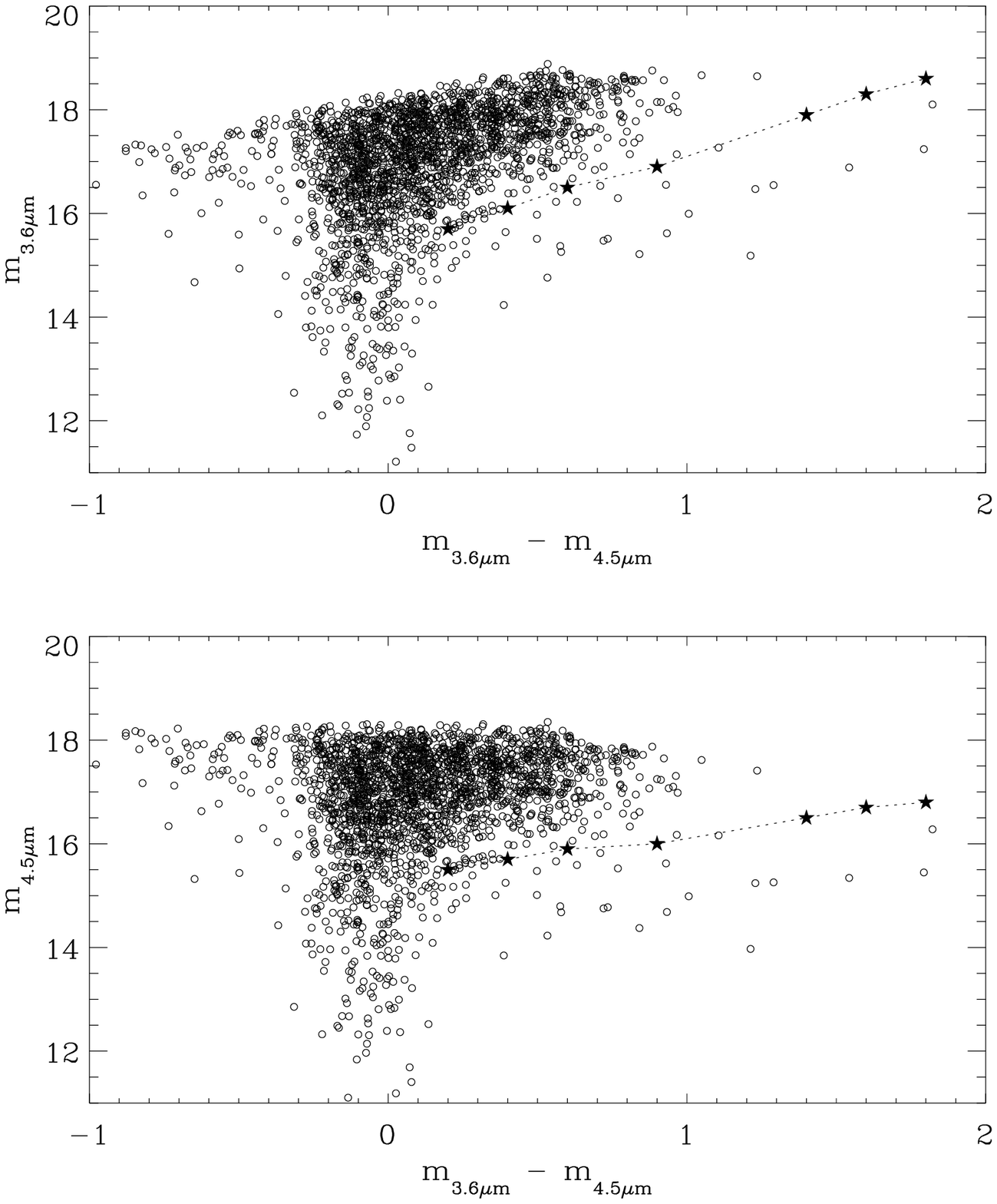}
\caption{IRAC 3.6 and 4.5 $\mu$m color-magnitude diagrams for all point-like sources in the 
fields of the 6 Hyades white dwarfs observed at both short wavelengths.  Also plotted ({\em stars}) 
are representative points in the IRAC T dwarf sequence (T1$-$T8; \citealt*{pat06}) at the 46 pc 
distance to the open cluster.
\label{fig10}}
\end{figure}

Figure \ref{fig10} plots color-magnitude diagrams for all detected point-like sources in the fields 
of the 6 Hyades white dwarfs observed in both of the short wavelength IRAC channels.  Included in
the plot is the expected T dwarf sequence (T1$-$T8; \citealt*{pat06}) at the 46 pc distance to the open 
cluster.  The cooler part of the T dwarf sequence appears to standout from most field objects.  All 
sources with $m_{3.6\mu{\rm m}}-m_{4.5\mu{\rm m}}>1.0$ were investigated individually, revealing 
a few extragalactic sources, spurious detections near bright stars or the edge of the mosaic, and 
sources with large photometric errors.  No reliable candidates near the T dwarf sequence were 
identified.  Since T and sub-T dwarfs should not be detected at the distance to the Hyades in the 
2 long wavelength IRAC channels, this limits what can be done with single epoch IRAC data to 
constrain wide, methane-bearing substellar companions.  Moreover, because very cool brown dwarfs 
will have $K-m_{3.6\mu{\rm m}}\ga1$ \citep*{pat06}, they should not be detected in 2MASS, as any such 
companions in the Hyades would have $K\ga16$ mag.  Therefore, further analysis of these datasets can 
only be be achieved with deep near-infrared imaging from the ground or proper motion analysis with a 
second epoch IRAC observation.

\section{LIMITS ON MASSIVE PLANETS AND COLD BROWN DWARF COMPANIONS}

Clearly, none of the 32 stars in Figures \ref{fig1}--\ref{fig8} display reliably measured photometric 
excess at 4.5 $\mu$m.  When combined with similar results for the white dwarfs analyzed in Paper I, 
there are a total of 47 observed degenerates which reveal no evidence for unresolved, cold brown 
dwarf or massive planetary companions.  This is a striking result, especially considering the large 
number of relatively young white dwarfs where massive planets would be detectable. There are 34 
white dwarfs for which the IRAC 4.5 $\mu$m observations were sensitive to unresolved planetary 
mass companions in the range $2-13$ $M_{\rm J}$, and 10 white dwarfs for which the data
were sensitive to brown dwarf companions in the range $14-20$ $M_{\rm J}$, according to 
substellar cooling models used here (I. Baraffe 2007, private communication; \citealt*{bar03}).

\subsection{The Influence of Stellar Evolution on Massive Planets}

The aperture photometry and image analysis places limits on the presence of unresolved or
partially resolved companions out to approximately 5 pixels or $6''$, the largest aperture used for 
photometry (Paper I).  This angle on the sky corresponds to a several hundred AU region around 
the target white dwarfs, which lie at typical distances in the range $d\approx20-100$ pc.  However, 
any planets which formed in the $5-50$ AU range during the main sequence phase, should now 
be located further out due to orbital expansion via mass loss during the asymptotic giant branch 
\citep*{far06,far05b,bur02,zuc87,jea24}, yet still within a region to which the IRAC photometry is 
sensitive.  While a typical expansion factor (equal to the ratio of the main sequence progenitor mass 
to the white dwarf mass) is $2-3$ for F-type stars, the massive young degenerates studied here are 
likely to produce much larger increases, up to a factor of 6 or so.

Tidal interactions should be relatively strong for massive planets orbiting intermediate mass 
stars, whose radii can become as large as several AU at the tip of the asymptotic giant branch.  
For example, a 5 $M_{\odot}$ main sequence star should grow to a maximum radius near 5 AU 
\citep*{vil07}, directly engulfing any planets in that range.  Using Equation 6 of \citet*{ras96} for such 
a star, and assuming a convective envelope mass near 3 $M_{\odot}$, a 10 $M_{\rm J}$ planet 
at 10 AU should tidally decay within the 1 Myr lifetime of its asymptotic giant branch host \citep*{vas93}.  
Although this calculation is likely oversimplified, it is instructive; eschewing direct engulfment is not 
sufficient for a planet to survive the post-main sequence.  Rather, the more massive the planet, the 
more it will induce tides in the giant star and the more likely it will experience orbital decay and be 
destroyed \citep*{ras96}.  One competing factor is the fact that by the time an asymptotic giant has 
reached it maximum radius there has been significant mass lost and the orbits of any planets should 
already have expanded commensurately.  However, a complete, time-dependent treatment of these 
competing forces -- orbital expansion due to mass loss versus orbital decay due to tidal forces -- on 
planets during the asymptotic giant branch has not been carried out.  Such a study would be highly
valuable but is beyond the scope of this paper.

\subsection{Massive Planets and Brown Dwarfs at Evolved Stars}

Recent work indicates a substantial percentage of substellar, radial velocity companions to 
first ascent giant stars are potentially or certainly above the deuterium burning minimum mass.
Presently 4 of 13 or 31\% of known substellar companions to giant stars have masses above 
13 $M_{\rm J}$ \citep*{liu07,lov07,nie07,hat05}.  Two items of interest for white dwarf planetary
system studies emerge from these results at giant stars.  First, the $M\sin{i}$ distribution of close 
substellar companions to intermediate mass, evolved stars is markedly different than for solar-type 
main sequence stars.  Second is the fact that all 4 brown dwarf hosting giants have 2 $M_{\odot} 
< M < 4$ $M_{\odot}$, and are thus related to the population of white dwarfs studied in this paper.

With orbital semimajor axes 0.5 AU $< a < 2.5$ AU \citep*{liu07,lov07,nie07}, all of the known 
substellar radial velocity companions to first ascent giant stars risk destruction during the ensuing 
asymptotic giant phase.  However, the most apparently brown dwarf-like companions have 
$M\sin{i}\approx20$ $M_{\rm J}$ \citep*{liu07,lov07}; potentially massive enough to eject the giant 
stellar envelope as have the $50-60$ $M_{\rm J}$ substellar companions to the white dwarfs WD 
0137$-$049 and GD1400 \citep*{bur06,max06,far05b,far04b}).  The close, $P\approx2$ hr substellar 
companion to WD 0137$-$049 is thought to have ejected the dense first ascent giant envelope of its 
host, as evidence by the fact the white dwarf is a low mass, helium core degenerate.  This is not the 
case for GD 1400B, whose degenerate host is a typical, carbon-oxygen core white dwarf, but its 
$P\approx10$ hr period (Burleigh et al. 2008, in preparation) indicates likely prior orbital decay 
due to the ejection of the asymptotic giant envelope.  By inference, both these substellar survivors 
would have originally orbited within roughly 1 AU of their host intermediate mass stellar progenitors.

\subsection{Formation, Persistence, and Sacrifice}

The results of this IRAC white dwarf study may imply that close massive planetary and brown 
dwarf companions to intermediate mass stars do not typically survive the asymptotic giant branch.  
Robust statistics are not yet available, but \citet*{lov07} estimate at least 3\% of stars with $M\ga1.8$ 
$M_{\odot}$ mass stars host $M\sin{i}>5$ $M_{\rm J}$ companions, and as discussed in \S5.2, about 
$1/3$ of these are brown dwarfs (according to the IAU definition).  Hence, a reasonably optimistic 
expectation for an IRAC search of 46 white dwarfs would be 1 or 2 detections.  This negative result 
suggests the possibility that a higher (substellar) companion mass is required to survive the entire 
post-main sequence -- specifically the asymptotic giant phase -- at orbital separations of a few to 
several AU.

However, recent evidence has suggested a mechanism which may allow planets to survive within 
the inner regions through both giant phases; sacrifice of the innermost components.  Analogous 
to the formation of several types of post-main sequence binaries, substellar companions can 
dynamically eject the (first or second phase) giant envelope \citep*{nel98,sok98}, thereby shielding 
any remaining components in a planetary system.  This process -- common envelope evolution -- is 
responsible for the close orbits of numerous low mass stellar and substellar companions to white 
dwarfs \citep*{far06,sch03}.  Essentially, unstable mass transfer from the giant results in a frictional 
exchange of orbital energy between the secondary and the envelope material, resulting in an orbit 
decrease and an efficient envelope ejection \citep*{pac76}.  The $M\sin{i}\approx3$ $M_{\rm J}$,
$a=1.7$ AU planet recently detected at the sdB star V391 Pegasi \citep*{sil07} has survived a first 
ascent giant phase involving significant mass loss, possibly due to such an interaction \citep*{han02,
nel98,sok98}.  Similarly, the $M\sin{i}\approx2$ $M_{\rm J}$, $a=2.5$ AU planet candidate at the 
pulsating white dwarf GD 66 \citep*{mul08}, may owe its inner region survival to the sacrfice of closer 
planets.

While intermediate separation planets may have indeterminate fates, any massive planets 
originally orbiting at $a\ga5$ AU, should now be located at tens to hundreds of AU in the white 
dwarf phase, having essentially eluded any substantial post-main sequence interaction with their 
host star \citep*{vil07}.  The results of this IRAC search imply that massive, $M\ga10$ $M_{\rm J}$ 
planets and brown dwarfs form rarely ($f\la3$\%) at these wide separations.

\subsection{Planets in the Hyades}

The massive planet recently found at the Hyades giant $\epsilon$ Tauri \citep*{sat07} is very interesting 
in light of the 7 classical white dwarf Hyads surveyed with IRAC -- would such an planet have been detected 
if it persisted into the white dwarf evolutionary phase?  The most likely mass of the planet at $\epsilon$ Tauri 
is near 10 $M_{\rm J}$, and the IRAC 4.5 $\mu$m photometry was sensitive to objects of this mass at 
virtually every Hyades target, although in reality such a detection might prove more or less difficult due to 
differences from the model predictions used here.  However, it is uncertain whether this Hyades planet at
$a=1.9$ AU will survive the asymptotic giant phase of its host, since the maximum radius of the star will
reach at least 3 AU \citep*{vil07}.

A clear advantage of the IRAC search of the Hyades white dwarfs is insensivitiy to orbital separation 
or inclination, parameter spaces which limit radial velocity and direct imaging searches for planets.  
The resulting substellar mass sensitivities achieved here for the Hyades are comparable to those produced 
via direct imaging with {\em HST} / NICMOS -- also around 10 $M_{\rm J}$ \citep*{fri05} -- but for {\em any} 
orbits out to $a\approx250$ AU.  Only very widely separated massive planets should have escaped detection.

\subsection{Cold Brown Dwarfs}

The previous sections have covered discussions of mass, but not of temperature.  For 27 targets 
in the survey, the IRAC 4.5 $\mu$m photometry was sensitive to the entire known T dwarf sequence, 
independent of the corresponding masses, and in a few cases more than a full magnitude fainter than 
a T8 dwarf \citep*{pat06}.  In fact, with the exception of a single target (G21-16; see \S5.7) whose IRAC 
image was confused with other sources, this survey rules out brown dwarf companions down to 25 
$M_{\rm J}$ within a few hundred AU of all white dwarf targets, implying a brown dwarf companion 
fraction less than around 2\%.  If the minimal sensitivity achieved here for these 46 targets is matched 
by similar IRAC 4.5 $\mu$m results for 124 nearby white dwarfs \citep*{mul07}, this fraction could be 
smaller than 0.6\%, and consistent with previous white dwarf studies which were sensitive to somewhat 
higher temperatures and masses (L dwarfs; \citealt*{hoa07,far05a,far04a,wac03,zuc87}).

\subsection{No Evidence for Merger Products}

Although not the focus of this study, the results give no indication of lasting merger products at any 
of the 10 additional massive white dwarfs observed specifically for this study.  Combined with the 
results of \citet*{han06}, there is as yet no indication of disks (and subsequent pollution) or reforged 
planets at nearly 2 dozen massive white dwarfs which could conceivably have formed as a merger 
of lower mass degenerates \citep*{lie05a}.  Given the recent results on externally-polluted white 
dwarfs with debris disks, it is possible that any dust disks formed via white dwarf mergers would dissipate 
rapidly, the particles becoming gaseous through mutual collisions within the disk (Paper I).

\subsection{Individual Objects}

{\em 0046$+$051} (vMa 2) At 4.4 pc, this degenerate represents a unique and advantageous hunting ground 
for planets and planetary system remnants.  This cool, helium atmosphere, metal-rich white dwarf has been 
externally polluted by interstellar or circumstellar matter.  Previous ground- and space-based mid-infrared 
imaging and photometry have ruled out the presence of a substellar companion suggested by \citet*{mak04}, 
down to $T_{\rm eff}\ga500$ K and corresponding to around 25 $M_{\rm J}$ at 5 Gyr \citep*{far04c,bur03}.
The IRAC 4.5 $\mu$m photometry of vMa 2 and the models used here rule out the presence of a companion 
as cool as 400 K, and a mass close to the deuterium burning limit at 4.4 Gyr.  Furthermore, the IRAC 4 channel 
color-color search for resolved substellar companions rules out the presence of any widely bound object as 
cool as $T_{\rm eff}\approx550$ K within $r\approx1200$ AU of vMa 2.  Models predict this should correspond 
to a mass of 25 $M_{\rm J}$ at the age of this well studied white dwarf.  

Deep ground-based $J$-band imaging observations have ruled out widely-bound planetary mass companions
to vMa 2 as small as 7 $M_{\rm J}$ at orbital separations near $10-200$ AU \citep*{bur08}.  There exist 2 epochs 
of IRAC 4.5 $\mu$m imaging of vMa 2 in the {\em Spitzer} archive, separated by 2.1 yr and clearly revealing $6.2''$ 
of proper motion upon blinking the aligned frames.  There are no field objects comoving with the white dwarf, ruling 
out well detected, resolved objects with $m_{4.5\mu{\rm m}}=16.7$ mag (see Figure \ref{fig9}) or 36 $\mu$Jy as 
companions within 1200 AU of vMa 2.  At the 4.4 pc distance and 4.4 Gyr age of vMa2, models predict that this 
very sensitive observational limit corresponds to a mass of 4 $M_{\rm J}$ and a temperature of just $T_{\rm eff}
\approx200$ K; signficantly lower than the known T dwarfs and only 40 K warmer than Jupiter(!).

{\em 0236$+$498} (EUVE J0239$+$50.0) Very little reliable photometry exists on this hot and faint 
degenerate, making it difficult to properly calibrate its spectral energy distribution (see Figure \ref{fig2}).  
The 2MASS $J$-band flux is likely the most reliable photometric data available, while the $H$-band flux 
has a large associated error.  The white dwarf is clearly detected in the short wavelength IRAC images, 
but significant uncertainty exists at the long wavelengths, where the flux in the photometric aperture could 
be due to a background source or noise.  Unfortunately, complications plague the short wavelength IRAC 
observations; a column pull-down artifact at the position of the white dwarf and a photometrically overlapping 
point source located $3.6''$ away.  The IRAF routine {\sf daophot} was used to deconvolve the flux of the white 
dwarf and the nearby point source, after manually correcting the column pull-down artifact by adding the median 
value of nearby columns.  Despite these efforts, it is possible or even likely the photometry of the white dwarf is
contaminated.  If the 2MASS $H$-band flux and IRAC photometry are somewhat accurate, then the white dwarf 
may possess an excess.  Ground-based near-infrared photometry is needed for this source.

{\em 0325$-$857AB}  (LB 9802AB) This visual pair has long been a binary suspect \citep*{han06,bar95} 
but its physical companionship is confirmed here for the first time.  Previous arguments for binarity relied 
on space density, which is insufficient to firmly establish companionship, especially in a rare system 
such as this, where the more massive of the pair white dwarfs is the hotter and apparently younger 
star.  Common proper motion is demonstrated by measuring the astrometric offsets of the white dwarfs 
and 20 background point sources between 2 epochs of SuperCOSMOS data.  Two scans of UKST 
photographic plates were used for this purpose, one image taken in 1976.7 and another from 2001.9, 
providing a 25.3 yr baseline.  Using the IRAF routine {\sf geomap} to measure both the motion of the
white dwarfs and the random centroiding errors from the positions of the background sources, the
component proper motions are: $\mu_A=0.074''$ ${\rm yr}^{-1}$ at $\theta_A=244\arcdeg$ and
$\mu_B=0.083''$ ${\rm yr}^{-1}$ at $\theta_B=242\arcdeg$, with $\sigma_{\mu}=0.012''$ ${\rm yr}^{-1}$ 
and $\sigma_{\theta}=6\arcdeg$.  The separation between primary and secondary from the 2MASS 
$J$-band image is $6.9''$ at position angle of $325\arcdeg$.

{\em 1455$+$298} (G166-58) This metal-rich white dwarf displays excess flux in the 2 long wavelength 
IRAC channels, but not at 4.5 $\mu$m (Paper I).  Therefore, it was included in the photometric analysis 
for unresolved substellar companions.

{\em 2326$+$049} (G29-38) Not included in the analysis for unresolved companions due to its large
photometric excess at all IRAC wavelengths (Paper I).  However, the field surrounding G29-38 
was searched for widely-bound, cold substellar companions using the method described in \S4.3, 
revealing no candidates down to 25 $M_{\rm J}$ within 3500 AU.

\section{CONCLUSION}

A relatively comprehensive look at both young and older white dwarfs with IRAC reveals no 
promising evidence for massive planets or cold brown dwarf companions at orbital separations
within a few hundred AU.  By conducting a search for substellar companions via excess emission 
in the mid-infrared, the data cover the widest range of orbital phase space possible; nearly all
possible semimajor axes, eccentricities and inclinations, including the missed middle ground
($5-50$ AU) between radial velocity and direct imaging searches.  The uncovered range of
possible orbits lies roughly beyond a few hundred AU.

While it is somewhat uncertain how substellar objects evolve (dynamically or otherwise) within $5-10$ 
AU of mass losing asymptotic giant stars, avoidance of direct engulfment may not suffice for ultimate 
survival to the white dwarf phase.  However, massive inner planets could act as sacrificial guardians for 
remaining outer planetary system components.   Any planets and brown dwarfs outside of this region should 
be relatively unaffected, and the IRAC results demonstrate a dearth of cold substellar companions down 
into sub-T dwarf temperatures.  The $2-3$ $M_{\rm J}$ planets suspected at the sdB star V 391 Pegasi 
and the white dwarf GD 66 are just beyond the reach of this survey.

The negative results for the Hyades and comparably aged white dwarfs yield upper mass limits at or 
somewhat below 10 $M_{\rm J}$ for objects which may have formed around 3 $M_{\odot}$ main sequence 
stars.  Similar limits at the Pleiades and analogously hot and massive field white dwarfs provide the first 
evidence that massive planets are not commonly formed or do not survive the post-main sequence evolution 
of intermediate mass stars with $M\ga4$ $M_{\odot}$.  This latter result was only achievable via observations
of white dwarfs, as their main sequence, B-type progenitors are not amenable to other planet detection 
techniques.

The lack of 4.5 $\mu$m excess at all white dwarf targets, especially when combined with similar 
IRAC searches \citep*{mul07}, confirms that L and T-type brown dwarf companions are rare ($f<0.6\%$) 
within a few hundred AU, down to masses near the deuterium burning limit.  These results suggest the
possibility that the lowest mass companions, and especially planets, orbiting intermediate mass stars 
may be altered or destroyed prior to or during the post-main sequence, or are (more likely) too cold to 
directly detect with current facilities.

\acknowledgments

The authors are grateful to referee F. Mullally for comments which resulted in an improved manuscript.
J. Farihi thanks M. Jura for helpful discussions on post-main sequence evolution.  This work is based 
on observations made with the {\em Spitzer Space Telescope}, which is operated by the Jet Propulsion 
Laboratory, California Institute of Technology under a contract with NASA.  Support for this work was 
provided by NASA through an award issued by JPL/Caltech, and by NASA grants to UCLA.

{\em Facility:} \facility{Spitzer (IRAC)}

\clearpage

\begin{deluxetable}{cccccc}
%\rotate
\tabletypesize{\footnotesize}
\tablecaption{Young Cluster White Dwarf Targets\label{tbl1}}
\tablewidth{0pt}
\tablehead{
\colhead{WD\#}			&
\colhead{Name}			&
\colhead{$T_{\rm eff}$ (K)}	&
\colhead{$V$ (mag)}			&
\colhead{$M$ ($M_{\odot}$)}	&
\colhead{References}}

\startdata

0349$+$247	&EG 25		&31700	&16.64	&1.09	&1,2,3\\
0352$+$096	&EG 26		&14800	&14.52	&0.71	&1,2,4\\
0406$+$169	&EG 29		&15200	&15.35	&0.80	&1,2,4\\
0415$+$271\tablenotemark{a}	
			&EG 265		&11500	&15.00	&0.55	&5,6,7\\
0421$+$162	&EG 36		&19600	&14.29	&0.68	&1,2,4\\
0425$+$168	&EG 37		&24400	&14.02	&0.70	&1,2,4\\
0431$+$126	&EG 39		&21300	&14.24	&0.65	&1,2,3\\
0437$+$138	&GR 316		&15300	&14.93	&0.68	&1,2,7\\
0438$+$108	&EG 42		&27400	&13.86	&0.75	&1,2,3\\	

\enddata

\tablenotetext{a}{EG 265 is a member of the Hyades cluster or 
supercluster \citep*{rei93,egg84}.}

\tablerefs{
(1) \citealt*{cla01};
(2) \citealt*{ber95a}; 
(3) \citealt*{che93};
(4) \citealt*{egg65};
(5) \citealt*{ber04};
(6) \citealt*{rei93};
(7) \citealt*{upg85}}

\end{deluxetable}

\clearpage

\begin{deluxetable}{cccccc}
%\rotate
\tabletypesize{\footnotesize}
\tablecaption{Young Field White Dwarf Targets\label{tbl2}}
\tablewidth{0pt}
\tablehead{
\colhead{WD\#}			&
\colhead{Name}			&
\colhead{$T_{\rm eff}$ (K)}	&
\colhead{$V$ (mag)}			&
\colhead{$M$ ($M_{\odot}$)}	&
\colhead{References}}

\startdata

0001$+$433	&EUVE		&42400	&16.8	&1.37		&1,2,3\\
0136$+$251	&PG			&39400	&15.87	&1.32		&1,3,4\\
0235$-$125	&PHL 1400	&32400	&14.98	&1.03		&1,2,5\\
0236$+$498	&EUVE		&33800	&15.8	&0.94		&6,7\\
0325$-$857A	&LB 9802A	&16200	&14.11	&0.85		&8,9\\
0325$-$857B	&LB 9802B	&33800	&14.90	&1.33		&1,8,10\\
0346$-$011	&GD 50		&43200	&14.04	&1.37		&1,2\\
0440$-$038	&EUVE		&65100	&16.7	&1.33		&1,2,7\\
0518$-$105	&EUVE		&33000	&15.82	&1.07		&1,2,3\\
0531$-$022	&EUVE		&29700	&16.20	&0.97		&6,7\\
0652$-$563	&EUVE		&35200	&16.6	&1.18		&1,2\\
0730$+$487	&GD 86		&15500	&14.96	&0.90		&12\\
0821$-$252	&EUVE 		&43200	&16.4	&1.21		&13\\
0914$-$195	&EUVE		&56400	&17.4	&1.33		&1,2\\
1022$-$301	&EUVE		&35700	&15.9	&1.27		&1,6,7\\
1440$+$753\tablenotemark{a}
			&EUVE 		&35000	&15.22	&1.06		&3,13,14\\
1529$-$772	&EUVE		&51600	&16.9	&1.24		&2,13\\
1543$-$366	&EUVE		&45200	&15.81	&1.17		&6,15\\
1609$+$631	&PG			&30400	&16.68	&1.05		&4,6\\
1642$+$413\tablenotemark{b}
			&PG			&26500	&16.21	&0.96		&4,6\\
1658$+$440	&PG			&30500	&15.02	&1.41		&1,4\\
1740$-$706	&EUVE		&46800	&16.51	&1.18		&1,13,15\\

\enddata

\tablenotetext{a}{1440$+$753 is a close double degenerate}

\tablenotetext{b}{\citet*{fin97} list 0.96 $M_{\odot}$ (109pc) for 1642$+$413, while \citet*{lie05a} 
give 0.79 $M_{\odot}$ (145 pc)}

\tablerefs{
(1) \citealt*{han06};
(2) \citealt*{ven97};
(3) \citealt*{mar97};
(4) \citealt*{lie05a};
(5) \citealt*{dup02};
(6) \citealt*{fin97};
(7) \citealt*{mcc06};
(8) \citealt*{bar95};
(9) \citealt*{fer97};
(10) \citealt*{ven03};
(12) \citealt*{ber92};
(13) \citealt*{ven99a};
(14) \citealt*{ven99b};
(15) \citealt*{ven96}}

\end{deluxetable}

\clearpage

\begin{deluxetable}{cccccc}
%\rotate
\tabletypesize{\footnotesize}
\tablecaption{Metal-Rich Field White Dwarf Targets\label{tbl3}}
\tablewidth{0pt}
\tablehead{
\colhead{WD\#}			&
\colhead{Name}			&
\colhead{$T_{\rm eff}$ (K)}	&
\colhead{$V$ (mag)}			&
\colhead{$M$ ($M_{\odot}$)}	&
\colhead{References}}

\startdata

0032$-$175	&G266-135	&9240	&14.94	&0.60	&1,2\\
0046$+$051	&vMa 2		&6770	&12.39	&0.83	&3\\
0235$+$064	&PG			&15000	&15.5	&0.61	&4\\
0322$-$019	&G77-50		&5220	&16.12	&0.61	&5,6\\
0846$+$346	&GD 96		&7370	&15.71	&0.59	&1,7\\
1102$-$183	&EC			&8060	&15.99	&0.60	&1,7\\
1124$-$293	&EC			&9680	&15.02	&0.63	&5,8\\
1204$-$136	&EC			&11500	&15.67	&0.60	&1,9\\
1208$+$576	&G197-47		&5880	&15.78	&0.56	&3\\
1344$+$106	&G63-54		&7110	&15.12	&0.65	&3\\
1407$+$425	&PG			&10010	&15.03	&0.73	&10\\
1455$+$298	&G167-8		&7390	&15.60	&0.58	&3,10\\
1632$+$177	&PG			&10100	&13.05	&0.58	&10\\
1633$+$433	&G180-63		&6690	&14.84	&0.72	&3,10\\
1826$-$045	&G21-16		&9480	&14.58	&0.57	&3\\
1858$+$393	&G205-52		&9470	&15.63	&0.60	&1,7\\
2326$+$049	&G29-38		&11600	&13.04	&0.69	&7,11\\

\enddata

\tablecomments{For those stars with no spectroscopic or trigonometric mass-radius estimate, 
log $g=8.0$ was assumed.}

\tablerefs{
(1) \citealt*{zuc03};
(2) \citealt*{mer86};
(3) \citealt*{ber01};
(4) This work;
(5) \citealt*{ber97};
(6) \citealt*{sma03};
(7) \citealt*{mcc06};
(8) \citealt*{koe01};
(9) \citealt*{sal03};
(10) \citealt*{lie05a};
(11) \citealt*{ber95d}}

\end{deluxetable}

\clearpage

\begin{deluxetable}{cccccc}
%\rotate
\tabletypesize{\footnotesize}
\tablecaption{IRAC Fluxes for White Dwarf Targets\label{tbl4}}
\tablewidth{0pt}
\tablehead{	
\colhead{WD\#}					&
\colhead{$F_{3.6\mu{\rm m}}$ ($\mu$Jy)}	&
\colhead{$F_{4.5\mu{\rm m}}$ ($\mu$Jy)}	&
\colhead{$F_{5.7\mu{\rm m}}$ ($\mu$Jy)}	&
\colhead{$F_{7.9\mu{\rm m}}$ ($\mu$Jy)}	&
\colhead{Pipeline}}

\startdata

0001$+$433	&$18\pm6$		&$13\pm7$		&30\tablenotemark{a}	&29\tablenotemark{a}	&14.0\\
0046$+$051	&$8040\pm400$	&$5360\pm270$	&$3680\pm190$		&$2080\pm110$		&14.0\\
0136$+$251	&$46\pm6$		&$26\pm7$		&34\tablenotemark{a}	&35\tablenotemark{a}	&14.0\\
0235$-$125	&$111\pm8$		&$63\pm7$		&36\tablenotemark{a}	&35\tablenotemark{a}	&14.0\\
0236$+$498	&$107\pm11$		&$72\pm9$		&20\tablenotemark{a}	&27\tablenotemark{a}	&11.0\\
0325$-$857A	&$396\pm20$		&$238\pm14$		&$139\pm27$			&$87\pm25$			&14.0\\
0325$-$857B	&$149\pm9$		&$92\pm8$		&$53\pm28$			&33\tablenotemark{a}	&14.0\\
0346$-$011	&$263\pm14$		&$164\pm11$		&$145\pm34$			&47\tablenotemark{a}	&14.0\\
0349$+$247	&$27\pm3$		&$15\pm4$		&22\tablenotemark{a}	&39\tablenotemark{a}	&11.4\\
0352$+$096	&$295\pm15$		&$168\pm9$		&$99\pm21$			&$44\pm28$			&11.4\\
0406$+$169	&$134\pm7$		&$69\pm5$		&$55\pm20$			&$61\pm34$			&11.4\\
0415$+$271	&$245\pm13$		&$148\pm8$		&$100\pm19$			&$68\pm23$			&11.4\\
0421$+$162	&$275\pm14$		&$163\pm9$		&$121\pm20$			&$58\pm23$			&11.4\\
0425$+$168	&$313\pm16$		&$185\pm10$		&$133\pm21$			&$101\pm27$			&11.4\\
0431$+$126	&\nodata			&$179\pm10$		&\nodata				&51\tablenotemark{a}	&14.0\\
0437$+$138	&$186\pm10$		&$107\pm6$		&$62\pm17$			&42\tablenotemark{a}	&11.4\\
0438$+$108	&\nodata			&$211\pm12$		&\nodata				&$76\pm38$			&14.0\\
0440$-$038	&$20\pm5$		&$19\pm6$		&23\tablenotemark{a}	&21\tablenotemark{a}	&14.0\\
0518$-$105	&$51\pm5$		&$33\pm7$		&32\tablenotemark{a}	&31\tablenotemark{a}	&14.0\\
0531$-$022	&$47\pm4$		&$26\pm5$		&20\tablenotemark{a}	&37\tablenotemark{a}	&11.4\\
0652$-$563	&$16\pm11$		&$18\pm9$		&32\tablenotemark{a}	&30\tablenotemark{a}	&14.0\\
0730$+$487	&$218\pm11$		&$127\pm7$		&$98\pm22$			&$66\pm24$			&10.5\\
0821$-$252	&$26\pm10$		&$13\pm7$		&21\tablenotemark{a}	&20\tablenotemark{a}	&11.0\\
0914$-$195	&$13\pm5$		&$9\pm6$			&33\tablenotemark{a}	&32\tablenotemark{a}	&14.0\\
1022$-$301	&$33\pm6$		&$22\pm7$		&30\tablenotemark{a}	&31\tablenotemark{a}	&14.0\\
1440$+$753	&$88\pm6$		&$58\pm5$		&$27\pm16$			&16\tablenotemark{a}	&11.0\\
1529$-$772	&$22\pm5$		&$12\pm4$		&20\tablenotemark{a}	&18\tablenotemark{a}	&12.4\\
1543$-$366	&$51\pm5$		&$31\pm5$		&23\tablenotemark{a}	&23\tablenotemark{a}	&11.4\\
1609$+$631	&$24\pm3$		&$13\pm3$		&15\tablenotemark{a}	&22\tablenotemark{a}	&11.0\\
1642$+$413	&$38\pm3$		&$21\pm4$		&16\tablenotemark{a}	&17\tablenotemark{a}	&11.4\\
1658$+$440	&$131\pm8$		&$81\pm7$		&$56\pm27$			&28\tablenotemark{a}	&14.0\\
1740$-$706	&$25\pm4$		&$14\pm4$		&16\tablenotemark{a}	&19\tablenotemark{a}	&14.0\\

\enddata

\tablenotetext{a}{Upper limit.}

\tablecomments{Photometric errors and upper limits are described in \S3.1.}

\end{deluxetable}

\clearpage

\begin{deluxetable}{ccccccc}
%\rotate
\tabletypesize{\footnotesize}
\tablecaption{Target Ages and Upper Mass Limits for Unresolved Companions\label{tbl5}}
\tablewidth{0pt}
\tablehead{
\colhead{WD\#}					&
\colhead{$t_{\rm ms}$ (Gyr)}			&
\colhead{$t_{\rm wd}$ (Gyr)}			&
\colhead{$\tau$\tablenotemark{a} (Gyr)}	&
\colhead{$d$ (pc)}					&
\colhead{$M_{4.5\mu{\rm m}}$ (mag)}	&
\colhead{Mass ($M_{\rm J}$)}}

\startdata

\cutinhead{Young Cluster White Dwarfs}

0349$+$247	&\nodata		&\nodata		&0.1		&132	&12.3		&9\\
0352$+$096	&\nodata		&\nodata		&0.6		&46		&13.7		&10\\
0406$+$169	&\nodata		&\nodata		&0.6		&46		&14.4		&7\\
0415$+$271	&\nodata		&\nodata		&0.6		&46		&13.9		&9\\
0421$+$162	&\nodata		&\nodata		&0.6		&46		&13.7		&10\\
0425$+$168	&\nodata		&\nodata		&0.6		&46		&13.6		&10\\
0431$+$126	&\nodata		&\nodata		&0.6		&46		&13.6		&10\\
0437$+$138	&\nodata		&\nodata		&0.6		&46		&14.2		&8\\
0438$+$108	&\nodata		&\nodata		&0.6		&46		&13.4		&11\\

\cutinhead{Young Field White Dwarfs}

0001$+$433	&\nodata		&\nodata		&0.1		&96		&12.4		&9\\
0136$+$251	&\nodata		&\nodata		&0.1		&73		&13.0		&6\\
0235$-$125	&\nodata		&\nodata		&0.1		&65		&13.3		&5\\
0236$+$498	&0.1			&0.1			&0.2		&107	&11.9		&11\\
0325$-$857A	&0.1			&0.3			&0.4		&35		&13.9		&6\\
0325$-$857B	&\nodata		&\nodata		&0.4		&35		&14.5		&5\\
0346$-$011	&\nodata		&\nodata		&0.1		&30		&14.5		&3\\
0440$-$038	&\nodata		&\nodata		&0.1		&134	&11.9		&10\\
0518$-$105	&\nodata		&\nodata		&0.1		&92		&12.5		&8\\
0531$-$022	&0.1			&0.1			&0.2		&107	&12.6		&9\\
0652$-$563	&\nodata		&\nodata		&0.1		&119	&11.7		&11\\
0730$+$487	&0.1			&0.4			&0.5		&39		&14.5		&6\\
0821$-$252	&\nodata		&\nodata		&0.1		&105	&12.2		&10\\
0914$-$195	&\nodata		&\nodata		&0.1		&175	&11.3		&12\\
1022$-$301	&\nodata		&\nodata		&0.1		&61		&13.4		&5\\
1440$+$753	&\nodata		&\nodata		&0.1		&101	&12.7		&7\\
1529$-$772	&\nodata		&\nodata		&0.1		&137	&12.3		&9\\
1543$-$366	&\nodata		&\nodata		&0.1		&111	&12.5		&8\\
1609$+$631	&\nodata		&\nodata		&0.1		&134	&12.6		&8\\
1642$+$413	&0.1			&0.1			&0.2		&109	&12.8		&8\\
1658$+$440	&\nodata		&\nodata		&0.1		&32		&14.8		&2\\
1740$-$706	&\nodata		&\nodata		&0.1		&76		&13.5		&5\\

\cutinhead{Metal-Rich Field White Dwarfs}

0032$-$175	&1.5			&0.7			&2.2		&31		&13.9		&18\\
0046$+$051	&0.2			&3.7			&3.9		&4.4		&15.2		&13\\
0235$+$064	&1.2			&0.2			&1.4		&70		&13.3		&20\\
0322$-$019	&1.2			&4.4			&5.6		&17		&15.1		&17\\
0846$+$346	&1.9			&1.4			&3.3		&30		&14.6		&17\\
1102$-$183	&1.5			&1.1			&2.6		&40		&14.2		&17\\
1124$-$293	&0.9			&0.6			&1.5		&34		&14.3		&12\\
1204$-$136	&1.5			&0.4			&1.9		&62		&13.5		&20\\
1208$+$576	&4.0			&2.5			&6.5		&20		&14.7		&25\\
1344$+$106	&0.7			&1.9			&2.6		&20		&14.7		&14\\
1407$+$425	&0.3			&0.9			&1.2		&33		&14.6		&10\\
1455$+$298	&2.4			&1.4			&3.8		&36		&14.1		&25\\
1632$+$177	&2.4			&0.6			&3.0		&16		&14.1		&20\\
1633$+$433	&0.3			&2.7			&3.0		&15		&14.8		&14\\
1826$-$045	&3.0			&0.8			&3.8		&29		&13.0		&35\\
1858$+$393	&1.5			&0.7			&2.2		&45		&14.2		&16\\

\enddata

\tablenotetext{a}{The total age of the Pleiad EG 25 (Table \ref{tbl1}) and the similarly hot and 
massive field white dwarfs (Table \ref{tbl2}) is taken to be 0.125 Gyr (see \S4.1).  The seventh
column lists the companion mass upper limit.}

\end{deluxetable}

\clearpage

\begin{deluxetable}{ccccccc}
%\rotate
\tabletypesize{\footnotesize}
\tablecaption{Target Ages and Upper Mass Limits for Resolved Companions\label{tbl6}}
\tablewidth{0pt}
\tablehead{
\colhead{WD\#}				&
\colhead{$t_{\rm ms}$ (Gyr)}		&
\colhead{$t_{\rm wd}$ (Gyr)}		&
\colhead{$\tau$ (Gyr)}			&
\colhead{$d$ (pc)}				&
\colhead{$M_{7.9\mu{\rm m}}$ (mag)}		&
\colhead{Mass ($M_{\rm J}$)}}

\startdata

0046$+$051	&0.2		&3.7		&3.9		&4.4		&16.0	&25\\
0322$-$019	&1.2		&4.4		&5.6		&17		&13.8	&65\\
1208$+$576	&4.0		&2.5		&6.5		&20		&13.5	&70\\
1344$+$106	&0.7		&1.9		&2.6		&20		&13.5	&60\\
1632$+$177	&2.4		&0.6		&3.0		&16		&14.0	&50\\
1633$+$433	&0.3		&2.7		&3.0		&15		&14.1	&50\\
2326$+$049	&0.4		&0.5		&0.9		&14		&14.3	&25\\

\enddata

\tablecomments{The exposure time for 0046$+$051 was 150 s versus 600 s for the other targets, 
resulting in an overall sensitivity about half of that calculated in \S4.3.1.}

\end{deluxetable}

\begin{deluxetable}{ccccc}
%\rotate
\tabletypesize{\footnotesize}
\tablecaption{Representative 4.5 $\mu$m Absolute Magnitudes for Substellar Objects\label{tbl7}}
\tablewidth{0pt}
\tablehead{
\colhead{Mass ($M_{\rm J}$)}	&
\colhead{0.1 Gyr}			&
\colhead{0.5 Gyr}			&
\colhead{1.0 Gyr}			&
\colhead{5.0 Gyr}}

\startdata

2	&14.8	&16.7	&17.7	&21.1\\
5	&13.2	&14.9	&15.7	&17.9\\
10	&12.0	&13.5	&14.3	&16.1\\
15	&11.1	&12.7	&13.5	&15.3\\
20	&10.9	&12.2	&12.9	&14.7\\

\enddata

\tablecomments{Entries are in magnitudes \citep*{ssc06a}.  The table is based on the models
of I. Baraffe (2007, private communication; \citealt*{bar03})}

\end{deluxetable}

\clearpage

\clearpage

\clearpage

\clearpage

\clearpage

\clearpage

\clearpage

\clearpage

\clearpage

\clearpage

\begin{thebibliography}{}

\bibitem[Baraffe et al.(2003)]{bar03} Baraffe, I., Chabrier, G., Barman, T. S., Allard, 
		F., \& Hauschildt, P. H. 2003, \aap, 402, 701

\bibitem[Barstow et al.(1995)]{bar95} Barstow, M. A., Jordan, S., O' Donoghue, 
		Burleigh, M. R., Napiwotzki, R., \& Harrop-Allin, M. K. 1995, \mnras, 277, 
		971
	
\bibitem[Barstow et al.(2005)]{bar05} Barstow, M. A., Bond, H. E., Holberg, J. B., 
		Burleigh, M. R., Hubeny, I., \& Koester, D. 2005, \mnras, 362, 1134	
		
\bibitem[Becklin, \& Zuckerman.(1988)]{bec88} Becklin, E. E., \& Zuckerman, B. 
		1988, \nat, 336, 656

\bibitem[Bergeron et al.(2004)]{ber04} Bergeron, P., Fontaine, G., Billeres, M., 
		Boudreault, S., \& Green, E. M. 2004, \apj, 600, 404

\bibitem[Bergeron et al.(2001)Bergeron, Leggett, \& Ruiz]	{ber01} Bergeron, P., 
		Leggett, S. K., \& Ruiz, M. T. 2001, \apjs, 133, 413

\bibitem[Bergeron et al.(1995a)Bergeron, Liebert, \& Fullbright]{ber95a} Bergeron, 
		P., Liebert, J., \& Fullbright, M. S. 1995a, \apj, 444, 810

\bibitem[Bergeron et al.(1997)Bergeron, Ruiz, \& Leggett]	{ber97} Bergeron, P., 
		Ruiz, M. T., \& Leggett, S. K. 1997, \apjs, 108, 339
		
\bibitem[Bergeron et al.(1992)Bergeron, Saffer, \& Liebert]{ber92} Bergeron, P., 
		Saffer, R. A., \& Liebert, J. 1992, \apj, 394, 228
	
\bibitem[Bergeron et al.(1995b)Bergeron, Saumon, \& Wesemael]{ber95b} Bergeron, 
		P., Saumon, D., \& Wesemael, F. 1995b, \apj, 443, 764

\bibitem[Bergeron et al.(1995c)Bergeron, Wesemael, \& Beauchamp]{ber95c} 
		Bergeron, P., Wesemael, F., \& Beauchamp, A. 1995c, \pasp, 107, 1047

\bibitem[Bergeron et al.(1995d)]{ber95d} Bergeron, P., Wesemael, F., Lamontagne, 
		R., Fontaine, G., Saffer, R. A., \& Allard, N. F. 1995d, \apj, 449, 258

\bibitem[Burleigh et al.(2002)Burleigh, Clarke, \& Hodgkin]{bur02} Burleigh, M. R., 
		Clarke, F. J., \& Hodgkin, S. T. 2002, \mnras, 331, L41
	
\bibitem[Burleigh et al.(2006)]{bur06} Burleigh, M. R., Hogan, E., Dobbie, P. D., 
		Napiwotzki, R., Maxted, P. F. L. 2006, MNRAS, 373, L55

\bibitem[Burleigh et al.(2008)]{bur08} Burleigh, M. R., et al. 2008, \mnras, 386, L5

\bibitem[Burrows et al.(2003)Burrows, Sudarsky, \& Lunine]{bur03} Burrows, A., 
		Sudarsky, D., \& Lunine, J. I. 2003, \apj, 596, 587

\bibitem[Carey(2006)]{car06} Carey, S. 2006, Spitzer Calibration Workshop, 
		(Pasadena: SSC)		

\bibitem[Chauvin et al.(2005)]{cha05} Chauvin, G., Lagrange, A. M., Dumas, C., 
		Zuckerman, B., Mouillet, D., Song, I., Beuzit, J. L., \& Lowrance, P. 2005,
		\aap, 438, L25	

\bibitem[Cheselka et al.(1993)]{che93} Cheselka, M., Holber, J., Watkins, R., 
		Collins, J., \& Tweedy, R. W. 1993, \aj, 106, 2365
	
\bibitem[Claver et al.(2001)]{cla01} Claver, C. F., Liebert, 	J., Bergeron, P., \& 
		Koester, D. 2001, \apj, 563, 987

\bibitem[Dahn et al.(2002)]{dah02} Dahn, C. C., et al. 2002, \aj, 124, 1170
		
\bibitem[Debes \& Sigurdsson(2002)]{deb02} Debes, J. H., \& 	Sigurdsson, S. 
		2002, \apj, 572, 556
		
\bibitem[Debes et al.(2007)Debes, Sigurdsson, \& Hansen]{deb07} Debes, J. H., 
		Sigurdsson, S., \& Hansen, B. 2007, \aj, 134, 1662
		
\bibitem[Dobbie et al.(2006a)]{dob06a} Dobbie, P. D., et al. 2006a, \mnras, 369, 383
	
\bibitem[Dobbie et al.(2006b)]{dob06b} Dobbie, P. D., Napiwotzki, R., Lodieu, N., 
		Burleigh, M. R., Barstow, M. A., \& Jameson, R. F. 2006b, \mnras, 373, L45	

\bibitem[Dupuis et al.(2002)Dupuis, Vennes, \& Chayer]{dup02} Dupuis, J., Vennes, 
		S., \& Chayer, P. 2002, \apj, 580, 1091

\bibitem[Eggen(1984)]{egg84} Eggen, O. J. 1984, \aj, 89, 830
		
\bibitem[Eggen \& Greenstein(1965)]{egg65} Eggen, O. J., \& Greenstein, J. L. 1965, 
		\apj, 141, 83
		
\bibitem[Farihi(2004)]{far04a} Farihi, J. 2004, Ph.D. Thesis, UCLA

\bibitem[Farihi et al.(2005a)Farihi, Becklin, \& Zuckerman]{far05a} Farihi, J., Becklin, 
		E. E., \& Zuckerman, B. 2005a, \apjs, 161, 394

\bibitem[Farihi et al.(2004)Farihi, Becklin, \& Macintosh]{far04c} Farihi, J., Becklin, 
		E. E., \& Macintosh, B. A. 2004, \apj, 608, L109

\bibitem[Farihi \& Christopher(2004)]{far04b} Farihi, J., \&	Christopher, M. 2004, \aj, 
		128, 1868

\bibitem[Farihi et al.(2006)Farihi, Hoard, \& Wachter]{far06} Farihi, J., Hoard, D. W., 
		\& Wachter, S. 2006, \apj, 646, 480

\bibitem[Farihi et al.(2005b)Farihi, Zuckerman, \& Becklin]{far05b} Farihi, J., Zuckerman, 
		B., \& Becklin, E. E. 2005b, \aj, 130, 2237
		
\bibitem[Farihi et al.(2008)Farihi, Zuckerman, \& Becklin]{far08} Farihi, J., Zuckerman, 
		B., \& Becklin, E. E. 2008, \apj, 674, 431 (Paper I)
	
\bibitem[Fazio et al.(2004)]{faz04} Fazio, G. G. et al.	2004, \apjs, 154, 10

\bibitem[Ferrario et al.(1997)]{fer97} Ferrario, L., Vennes,	S., Wickramasinghe, D. T., 
		Bailey, J. A., \& Christian, D. J. 1997, \mnras, 292, 205

\bibitem[Ferrario et al.(2005)]{fer05} Ferrario, L., Wickramasinghe, D. T., Liebert, J., 
		\& Williams, K. A. 2005, \mnras, 361, 1131

\bibitem[Finley et al.(1997)Finley, Koester, \& Basri]{fin97} Finley, D. S., Koester, D., 
		\& Basri, G. 1997, \apj, 488, 375

\bibitem[Friedrich et al.(2005)]{fri05} Friedrich, S., Zinnecker, H., Brandner, W., Correia, S.,
		\& McCaughrean, M. 2005, Proceedings of the	$14^{\rm th}$ European Workshop 
		on White Dwarfs, eds. D. Koester \& S. moehler (San Francisco: ASP), 431

\bibitem[Frink et al.(2002)]{fri02} Frink, S., Mitchell, D. S., Quirrenbach, A., Fischer, 
		D. A., Marcy, G. W., \& Butler, R. P. 2002, \apj, 576, 478
	
\bibitem[Glass(1999)]{gla99} Glass, I. S. 1999, Handbook of Infrared Astronomy, 
		(Cambridge; New York: Cambridge University Press)

\bibitem[Han et al.(2002)]{han02} Han, Z., Podsiadlowski, P., Maxted, P. F. L., Marsh, T. R.,
		\& Ivanova, N. 2002, \mnras, 336, 449

\bibitem[Hansen et al.(2006)Hansen, Kulkarni, \& Wiktorowicz]	{han06} Hansen, B. 
		M. S., Kulkarni, S., \& Wiktorowicz, S. 2006, \aj, 131, 1106

\bibitem[Hatzes et al.(2005)]{hat05} Hatzes, A. P., Guenther, E. W., Endl, M., Cochran, 
		W. D., D\"ollinger, M. P., \& Bedalov, A. 2005, \aap, 437, 743
	
\bibitem[Hatzes et al.(2006)]{hat06} Hatzes, A. P., et al. 2006, \aap, 457, 335

\bibitem[Hoard et al.(2007)]{hoa07} Hoard, D. W., Wachter, S., Sturch, L. K., Widhalm, 
		A. M., Weiler, K. P., Pretorius, M. L., Wellhouse, J. W., Gibiansky, M. 2007, 
		\aj, 134, 26

\bibitem[Hurley et al.(2000)Hurley, Pols, \& Tout]{hur00} Hurley, J. R., Pols, O. R., \& 
		Tout, C. A. 2000, \mnras, 315, 543

\bibitem[Jeans(1924)]{jea24} Jeans, J. 1924, \mnras, 85, 2

\bibitem[Jura(2003)]{jur03} Jura, M. 2003, \apj, 584, L91

\bibitem[Jura et al.(2007a)Jura, Farihi, \& Zuckerman]{jur07a} Jura, M., Farihi, J., \& 
		Zuckerman, B. 2007a, \apj, 663, 1285
	
\bibitem[Jura et al.(2007b)]{jur07b} Jura, M., Farihi, J., Zuckerman, B., \& Becklin, E. 
		E. 2007b, \aj, 133, 1927

\bibitem[Kalirai et al.(2008)]{kal08} Kalirai, J. S., Hansen, B. M. S., Kelson, D. D., 
		Reitzel, D. B., Rich, R. M., \& Richer, H. B. 2008, \apj, 676, 594

\bibitem[Kalirai et al.(2005)]{kal05} Kalirai, J. S., Richer, H. B., Reitzel, D., Hansen, 
		B. M. S., Rich, R. M., Fahlman, G. G., Gibson, B. K., \& von Hippel, T. 2005,
		\apj, 618, L123

\bibitem[Kilic \& Redfield(2007)]{kil07} Kilic, M., \& Redfield, S. 2007, \apj, 660, 641 

\bibitem[Kirkpatrick et al.(1999)]{kir99} Kirkpatrick, J. D., Allard, F., Bida, T., Zuckerman, 
		B., Becklin, E. E., Chabrier, G., \& Baraffe, I. 1999, \apj, 519, 834

\bibitem[Koester et al.(2001)]{koe01} Koester, D., et al. 2001, \aap, 378, 556
	
\bibitem[Liebert et al.(2005a)Liebert, Bergeron, \& Holberg]{lie05a} Liebert, J., 
		Bergeron, P., \& Holberg, J. B. 2005a, \apjs, 156, 47

\bibitem[Liebert et al.(2005b)]{lie05b} Liebert, J., Young, P. A., Arnett, D., Holberg, 
		J. B., \& Williams, K. A. 2005b, \apj, 630, L69

\bibitem[Liu et al.(2007)]{liu07} Liu, Y. J., et al. 2007, \apj, 672, 553 

\bibitem[Lovis \& Mayor(2007)]{lov07} Lovis, C., \& Mayor, M. 2007, \aap, 472, 657

\bibitem[Makarov(2004)]{mak04} Makarov, V. V. 2004, \apj, 600, L71

\bibitem[Marsh et al.(1997)]{mar97} Marsh, M. C., et al. 1997, \mnras, 286, 369
	
\bibitem[Maxted et al.(2006)]{max06} Maxted, P. F. L., Napiwotzki, R., Dobbie, P. D., 
		\& Burleigh, M. R. 2006, \nat, 442, 543

\bibitem[McCook \& Sion(2006)]{mcc06} McCook, G. P., \& Sion, E. M. 2006, 
		Spectroscopically Identified White Dwarfs (Strasbourg: CDS)

\bibitem[Mermilliod(1986)]{mer86} Mermilliod, J. C. 1986, Catalog of Eggen's 
		UBV Data (Strasbourg: CDS)
	
\bibitem[Mullally et al.(2007)]{mul07} Mullally, F., Kilic, M., Reach, W. T., Kuchner, 
		M. J., von Hippel, T., Burrows, A., \& Winget, D. E. 2007, \apjs, 171, 206

\bibitem[Mullally et al.(2008)]{mul08} Mullally, F., Winget, D. E., Degennaro, S., Jeffery, 
		E., Thompson, S. E., Chandler, D., \& and Kepler, S. O. 2008, \apj, 676, 573
	
\bibitem[Nakajima et al.(1995)]{nak95} Nakajima, T., Oppenheimer, B. R., Kulkarni, 
		S. R., Golimowski, D. A., Matthews, K., \& Durrance, S. T. 1995, \nat, 378, 
		463

\bibitem[Nelemans \& Tauris(1998)]{nel98} Nelemans, G., \& Tauris, T. M. 1998,
		\aap, 335, L85
	
\bibitem[Niedzielski et al.(2007)]{nie07} Niedzielski, A., et al. 2007, \apj, 669, 1354

\bibitem[Paczynski(1976)]{pac76} Paczynski, B. 1976, Proceedings of IAU Symposium 
		73, eds. P. Eggleton, S. Mitton, \& J. Whelan (Dordrecht: D. Reidel), 75	
\bibitem[Patten et al.(2006)]{pat06} Patten, B. M., et al. 2006, \apj, 651, 502

\bibitem[Perryman et al.(1998)]{per98} Perryman, M. A. C., et al. 1998, \aap, 331, 81

\bibitem[Pinsonneault et al.(1998)]{pin98} Pinsonneault, M. H., Stauffer, J., 
		Soderblom, D. R., King, J. R., \& Hanson, R. B. 1998, \apj, 504, 170
	
\bibitem[Probst(1983)]{pro83} Probst, R. 1983, \apjs, 53, 335

\bibitem[Rasio et al.(1996)]{ras96} Rasio, F. A., Tout, C. A., Lubow, S. H., \& Livio, M. 
		1996, \apj, 470, 1187
	
\bibitem[Reach et al.(2005)]{rea05} Reach, W. T., Kuchner, 	M. J., von Hippel, 
		T., Burrows, A., Mulally, F., Kilic, M., \& Winget, D. E. 2005a, \apj, 635, L161.
	
\bibitem[Reffert et al.(2006)]{ref06} Reffert, S., Quirrenbach, A., Mitchell, D. S., 
		Albrecht, S., Hekker, S., Fischer, D. A., Marcy, G. W., \& Butler, R. P. 
		2006, \apj, 652, 661

\bibitem[Reid(1993)]{rei93} Reid, I. N. 1993, \mnras, 265, 785

\bibitem[Salim \& Gould(2003)]{sal03} Salim, S., \& Gould, A. 2003, \apj, 582, 1011

\bibitem[Sato et al.(2003)]{sat03} Sato, B., et al. 2003, \apj, 597, L157

\bibitem[Sato et al.(2007)]{sat07} Sato, B., et al. 2007, \apj, 661, 527

\bibitem[Schreiber \& G\"ansicke(2003)]{sch03} Schreiber, M. R., \& G\"ansicke,
		B. T 2003, \aap, 406, 305
	
\bibitem[Silvotti et al.(2007)]{sil07} Silvotti, R., et al. 2007, \nat, 449, 189

\bibitem[Smart et al.(2003)]{sma03} Smart, R. L., et al. 2003, \aap, 404, 317
	
\bibitem[Soderblom et al.(1998)]{sod98} Soderblom, D. R., King, J. R., Hanson, R. 
		B., Jones, B. F., Fischer, D. A., Stauffer, J. R., \& Pinsonneault, M. H. 1998, 
		\apj, 504, 192
		
\bibitem[Soker(1998)]{sok98} Soker, N. 1998, \aj, 116, 1308

\bibitem[Song et al.(2006)]{son06} Song, I., Schneider, G., Zuckerman, B., Farihi, J., 
		Becklin, E. E., Bessell, M. S., Lowrance, P., \& Macintosh, B. A. 2006, \apj,
		652, 724

\bibitem[Spitzer Science Center(2006a)]{ssc06a} Spitzer Science Center. 2006a, 
		IRAC Data Handbook Version 3.0 (Pasadena: SSC)

\bibitem[Spitzer Science Center(2006b)]{ssc06b} Spitzer Science Center. 2006b, 
		Spitzer Observer's Manual Version 7.1 (Pasadena: SSC)	
	
\bibitem[Stauffer et al.(1998)Stauffer, Schultz, \& Kirkpatrick]{sta98} Stauffer, J. R., 
		Schultz, G., \& Kirkpatrick, J. D. 1998, \apj, 499, L199

\bibitem[Trembley \& Bergeron(2007)]{tre07} Tremblay, P. E., \& Bergeron, P. 2007, \apj, 
		657, 1013

\bibitem[Upgren et al.(1985)Upgren, Weis, \& Hanson]{upg85} Upgren, A. R., Weis, 
		E. W., \& Hanson, R. B. 1985, \aj, 90, 2039

\bibitem[Vassiliadis \& Wood(1993)]{vas93} Vassiliadis, E., \& Wood, P. R. 1993,
		\apj, 413, 641
			
\bibitem[Vennes(1999)]{ven99a} Vennes, S. 1999, \apj, 525, 995

\bibitem[Vennes et al.(1999)Vennes, Ferrario, \& Wickramasinghe]{ven99b} Vennes, S., 
		Ferrario, L., \& Wickramasinghe, D. T. 1999, 302, L99

\bibitem[Vennes et al.(2003)]{ven03} Vennes, S., Schmidt, G. D., Ferrario, L., Christian, 
		D. J., Wickramasinghe, D. T., \& Kawka, A. 2003, \apj, 593, 1040

\bibitem[Vennes et al.(1997)]{ven97} Vennes, S., Thejll, P. A., Galvan, R. G., \& Dupuis, 
		J. 1997, \apj, 480, 714
	
\bibitem[Vennes et al.(1996)]{ven96} Vennes, S., Thejll, P. A., Wickramasinghe, D. T., 
		Bessell, M. S. 1996, \apj, 467, 782	

\bibitem[Villaver \& Livio(2007)]{vil07} Villaver, E., \& Livio, M. 2007, \apj, 661, 1192
	
\bibitem[Wachter et al.(2003)]{wac03} Wachter, S., Hoard, D. W., Hansen, K. H., Wilcox, 
		R. E., Taylor, H. M., \& Finkelstein, S. L. 2003, \apj, 586, 1356

\bibitem[Weidemann(1987)]{wei87} Weidemann, V. 1987, \aap, 188, 74

\bibitem[Weidemann(1990)]{wei90} Weidemann, V. 1990, \aapr, 28, 103

\bibitem[Weidemann(2000)]{wei00} Weidemann, V. 2000, \aap, 363, 647

\bibitem[Werner et al.(2004)]{wer04} Werner, M. W., et al. 2004, \apjs, 154, 1

\bibitem[Williams et al.(2004)Williams, Bolte, \& Koester]{wil04} Williams, K. A., Bolte, 
		M., \& Koester, D. 2004, \apj, 615, L49

\bibitem[Wolszczan \& Frail(1992)]{wol92} Wolszczan, A., \& Frail, D. A. 1992, \nat, 
		355, 145
 	
\bibitem[Zuckerman \& Becklin(1987)]{zuc87} Zuckerman, B., \& Becklin, E. E. 1987, 
		\apj, 319, 99
		
\bibitem[Zuckerman et al.(2003)]{zuc03} Zuckerman, B., Koester, D., Reid, I. N., \& 
		H\"unsch, M. 2003, \apj, 596, 477
		
\end{thebibliography}
\end{document}